\input harvmac
\noblackbox

\def\tilde{\widetilde}
\newcount\figno
\figno=0
\def\fig#1#2#3{
\par\begingroup\parindent=0pt\leftskip=1cm\rightskip=1cm\parindent=0pt
\baselineskip=11pt
\global\advance\figno by 1
\midinsert
\epsfxsize=#3
\centerline{\epsfbox{#2}}
\vskip 12pt
\centerline{{\bf Figure \the\figno:} #1}\par
\endinsert\endgroup\par}
\def\figlabel#1{\xdef#1{\the\figno}}

\def\np#1#2#3{Nucl. Phys. {\bf B#1} (#2) #3}
\def\pl#1#2#3{Phys. Lett. {\bf B#1} (#2) #3}

\def\physrev#1#2#3{Phys. Rev. {\bf D#1} (#2) #3}

\def\cmp#1#2#3{Comm. Math. Phys. {\bf #1} (#2) #3}

\def\jmp#1#2#3{J.Math.Phys. {\bf #1} (#2) #3}
\def\ijmp#1#2#3{Int. J. of Modern Physics A, {\bf #1} (#2) #3}
\def\lmp#1#2#3{Let. Math. Phys. {\bf #1} (#2) #3}
\def\cqg#1#2#3{Class. Quantum Grav. {\bf #1} (#2) #3}


\font\cmss=cmss10
\font\cmsss=cmss10 at 7pt
\def\rlx{\relax\leavevmode}
\def\inbar{\vrule height1.5ex width.4pt depth0pt}
\def\IC{\relax\,\hbox{$\inbar\kern-.3em{\rm C}$}}
\def\IN{\relax{\rm I\kern-.18em N}}
\def\IP{\relax{\rm I\kern-.18em P}}
\def\ZZ{\rlx\leavevmode\ifmmode\mathchoice{\hbox{\cmss Z\kern-.4em Z}}
 {\hbox{\cmss Z\kern-.4em Z}}{\lower.9pt\hbox{\cmsss Z\kern-.36em Z}}
 {\lower1.2pt\hbox{\cmsss Z\kern-.36em Z}}\else{\cmss Z\kern-.4em
 Z}\fi} 
\def\IZ{\relax\ifmmode\mathchoice
{\hbox{\cmss Z\kern-.4em Z}}{\hbox{\cmss Z\kern-.4em Z}}
{\lower.9pt\hbox{\cmsss Z\kern-.4em Z}}
{\lower1.2pt\hbox{\cmsss Z\kern-.4em Z}}\else{\cmss Z\kern-.4em
Z}\fi}

\def\narrowplus{\kern -.04truein + \kern -.03truein}
\def\narrowminus{- \kern -.04truein}
\def\narrowminussub{\kern -.02truein - \kern -.01truein}

\def\half{{1\over 2}}

\def\cl{\centerline}

\def\b{{\beta}}
\def\a{{\alpha}}
\def\g{{\gamma}}
\def\ep{{\epsilon}}
\def\d{{\delta}}

\def\t{{\theta}}
\def\l{{\lambda}}

\def\s{{\sigma}}

\def\r{{\rightarrow}}

\def\lb{{\overline \lambda}}
\def\eb{{\overline \epsilon}}
\def\frac#1#2{{#1\over #2}}

\def\CA{{\cal A}}

\def\IZ{\relax\ifmmode\mathchoice
{\hbox{\cmss Z\kern-.4em Z}}{\hbox{\cmss Z\kern-.4em Z}}
{\lower.9pt\hbox{\cmsss Z\kern-.4em Z}}
{\lower1.2pt\hbox{\cmsss Z\kern-.4em Z}}\else{\cmss Z\kern-.4em
Z}\fi}
\def\IB{\relax{\rm I\kern-.18em B}}
\def\IC{{\relax\hbox{$\inbar\kern-.3em{\rm C}$}}}
\def\ID{\relax{\rm I\kern-.18em D}}
\def\IE{\relax{\rm I\kern-.18em E}}
\def\IF{\relax{\rm I\kern-.18em F}}
\def\IG{\relax\hbox{$\inbar\kern-.3em{\rm G}$}}
\def\IGa{\relax\hbox{${\rm I}\kern-.18em\Gamma$}}
\def\IH{\relax{\rm I\kern-.18em H}}
\def\II{\relax{\rm I\kern-.18em I}}
\def\IK{\relax{\rm I\kern-.18em K}}
\def\IP{\relax{\rm I\kern-.18em P}}

\def\p{\partial}

\font\cmss=cmss10 \font\cmsss=cmss10 at 7pt
\def\IR{\relax{\rm I\kern-.18em R}}

\def\e{{\epsilon_0}}
\def\Gt{ \tilde{\Gamma}}
\def\at{ \tilde{\alpha}}
\def\bt{ \tilde{\beta}}
\def\pslash{P\llap{/}}
\def\kslash{K\llap{/}}
%

%
%
\def\eqnn#1{\xdef #1{(\secsym\the\meqno)}\writedef{#1\leftbracket#1}%
\global\advance\meqno by1\wrlabeL#1}
\def\eqna#1{\xdef #1##1{\hbox{$(\secsym\the\meqno##1)$}}
\writedef{#1\numbersign1\leftbracket#1{\numbersign1}}%
\global\advance\meqno by1\wrlabeL{#1$\{\}$}}
\def\eqn#1#2{\xdef #1{(\secsym\the\meqno)}\writedef{#1\leftbracket#1}%
\global\advance\meqno by1$$#2\eqno#1\eqlabeL#1$$}

\lref\revans{N.T.Evans, ``Discrete Series for the Universal Covering Group of
the 3+2 dimensional de Sitter Group'',\jmp{8}{1967}{170-185}.}
\lref\rmack{G. Mack, ``All Unitary Ray Representations of the Conformal 
Group SU(2,2) with Positive Energy,\cmp{55}{1977}{1}''}
\lref\rdpresult{V.K.Dobrev and V.B. Petkova ``All Positive Energy Unitary Irreducible Representations Of Extended Conformal Supersymmetry,''
\pl{162B}{1985}{127-132}}
\lref\rdpdetails{V.K.Dobrev and V.B. Petkova, ``Group Theoretic Approach To 
Extended Conformal Supersymmetry...,'' Fortschr. Phys. {\bf 35} (1987) 537-572,
.The relevant portions are contained mainly in section 7 }
\lref\rdpreview{V.K.Dobrev and V.B. Petkova, ``All Positive Energy Unitary Irreducible Representations Of Extended Conformal Supersymmetry,'' in `Conformal 
Groups and Related Symmetries Physical Results and Mathematical Background, 
Springer Verlag Lecture notes in Physics 261,  Proceedings of a Symposium at
Clausthal, Eds A.O.Barut and H.D. Doebner}
\lref\rkacmain{V.G.Kac, ``Representations of Classical Lie Superalgebras'', in 
Lecture Notes in Mathematics, Vol. 676 (Springer-Verlag, Berlin, 1978) 597 }
\lref\rkacreview{V.G.Kac, ``Lie Superalgebras'', Advances in Mathematics, 
{\bf 26} (1977), 8-96.}
\lref\rshnider{Steven Shnider, ``The Superconformal Algebra in Higher 
Dimensions ''\lmp{16}{1988}{377-383}.}
\lref\rnahm{M. Scheunert, W.Nahm and V. Rittenberg, ``Classification of all 
simple graded Lie algebras whose Lie algebra is reductive. 2. Construction of 
the exceptional algebras''\jmp{17}{1976}{1640-1644}, see appendix C.} 
\lref\rheidenreich{W. Heidenreich, ``All Linear Unitary Irreducible 
Representations of de Sitter Supersymmetry with Positive Energy,'' 
\pl{110}{1982}{461-464}.}
\lref\rfreedman{D.Z.Friedman and H.Nicolai,  ``Multiplet Shortening in Osp(N,4)
'', \np{237}{1984}{342-366}}
\lref\rstrathdee{J.Strathdee, ``Extended Poincare Supersymmetry'',
\ijmp{2}{1987}{273-300}}
\lref\rsiegel{W.Siegel, ``All Free Conformal representations in all Dimensions,
'' \ijmp{4}{1989}{2015-2020}}
\lref\rconf{ Two references on the rather strange fact that scale invariance 
often implies conformal invariance are - S.Coleman, `` Aspects of Symmetry'', 
Cambridge University Press, Lecture 3; and J. Polchinksi, `` Scale and 
Conformal invariance in Quantum Field Theory'', \np{303}{1988}{226-236} }
\lref\rseiberg{Nathan Seiberg, ``Notes on Theories with 16 Supercharges'', 
hep-th/9705117 }
\lref\rroomana{A.Casher, F.Englert, H. Nicolai , M.Rooman, 
``The Mass Spectrum of Supergravity on the round seven sphere'',
\np{243}{1984}{173-188} }
\lref\rgunyadina{M.Gunaydin and N.P. Warner, 
``Unitary Supermultiplets of OSp(8/4,R) ... Supergravity,'' \np{272}{1986}{99-124}}
\lref\rgunaydinb{M.Gunaydin and N. Marcus, 
``The Spectrum of the $S^5$ compactification of the Chira N=2, D=10... 
U(2,2/4),'' \cqg{2}{1985}{L11-L17}}
\lref\rgunaydinc{M.Gunaydin and P.  van Nieuwenhuizen, and N.P  Warner, 
``General construction of the unitary representaion of $AdS$ superalgebras
and the spectrum of the $S^7$ compactification of 11-dimensional Supergravity
'' \np{255}{1985}{63-92}}
\lref\rnieua{H.J.Kim, L.J. Romans and P.Nieuwenhuizen, 
``Mass Spectrum of chiral 10d N=2 supergravity on $S^5$ ,'' \physrev{32}{1985}
{389-399}}
\lref\rnieub{H.J.Kim, L.J. Romans and P.Nieuwenhuizen, 
``Mass Spectrum of chiral 10d N=2 supergravity on $S^5$ ,'' \physrev{32}{1985}
{389-399}}
\lref\rkallosh{P.Claus, R. Kallosh and A V Proeyen, 
``M-5 brane and superconformal(0,2) tensor multiplet in 6 dimensions ,''
hep-th/9711161}

\Title{ \vbox{\baselineskip12pt\hbox{hep-th/9712074}
\hbox{PUPT-1748}}}
{\vbox{\centerline{Restrictions imposed by Superconformal Invariance}
\centerline{On Quantum Field Theories}
}}

\smallskip
\centerline{Shiraz Minwalla\footnote{$^1$}{minwalla@princeton.edu}}
\medskip\centerline{\it The Department of Physics}
\centerline{\it Princeton University}
\centerline{\it Princeton, NJ 08540, USA}

\vskip 1in

\noindent 
We derive unitarity restrictions on the scaling dimensions of primary 
quantum operators in a superconformal quantum field theory, in d =3, 4, 5, 6.

\vskip 0.1in
\Date{12/97}

\newsec{Introduction}
	Whereas classical massless field theories are often conformally
invariant, the same is rarely true of their quantum counterparts. To
 define a quantum field theory, you need to specify both a 
Lagrangian and a cutoff. The cutoff introduces a hidden dimensional scale 
in the definition of the theory. A scale transformation relates a quantum 
theory with a given cutoff to another quantum theory with the scaled cutoff. 
Therefore a QFT with nontrivial cutoff dependence (i.e., nonzero beta function) 
is never scale invariant, and so never conformally invariant. 

	Since quantum field theories generically have nonzero beta function, 
a study of the implications of exact conformal invariance on a quantum field 
theory might seem to be of limited interest. That is, however, not the case.
A QFT is indeed not scale invariant at arbitrary cut off, but as you lower 
the cut off in your theory, a truly gapless theory flows, in the 
Wilsonian sense, to a fixed point of the renormalization 
group - i.e., a point with zero beta function. Long wavelength physics
in such a theory is thus governed by an effective quantum theory that is 
exactly scale invariant. In many examples this effective quantum field theory
is a free massless theory. In some cases of interest however, the low energy effective 
theory is an interacting scale invariant, and 
generically conformally invariant \rconf\ quantum field theory.
 Low energy
effective theories completely specify the vacuum structure of a theory, 
and the nature of its massless excitations, and hence are of interest.

	Supersymmetric field theories have proved to be particularly amenable
to exact analysis in the infrared. Interacting gapless theories with unbroken 
supersymmetry typically exhibit both conformal and supersymmetric invariance in
the infrared. The full symmetry of the (low energy effective) 
theory must then be generated by a superalgebra that contains in it, as sub 
super algebras, both the conformal algebra and the supersymmetry algebra. 
Such algebras are very constrained - in fact none exist in $d \ge$6, and their 
possible forms are known in $d \leq 6$. These algebras are called superconformal
 algebras, and the symmetry they generate is called superconformal symmetry.

	Three examples of interacting superconformal theories of current 
interest are the intrinsic theories living on the world volumes of coincident 
$M_2, M_5 , D_3$ branes. A fourth example is the theory that governs the low
energy behaviour of $N=1, d=4$ super QCD with $N_f$ quarks and $N_c$ colours
with ${3 \over 2}N_c < N_f <3N_c $.

	In this paper we will derive unitarity constraints on quantum field 
theories in $d$ = 3, 4, 5, 6 that exactly implement the superconformal algebra.
The constraints we will derive will take the form of inequalities that scaling
dimensions of primary local operators in such a theory will be forced to obey. 

	This paper is arranged as follows.  

In Section 2 we will determine constraints that conformal invariance imposes
on a quantum theory in arbitrary dimension. 

	In section 3 we will construct the superconformal algebras that exist,
i.e. list all commutation relations explicitly.
We will also determine some constraints on scaling dimension, 
imposed by demanding the unitarity of these representations.

	In section 4 we will review  well known results in the theory of 
Lie Super Algebras, following Kac \rkacreview. In particular we will review 
the classification of such algebras, and define explicitly all algebras that 
we will use in the rest of the paper. We will explicitly, following \rshnider
, identify the superalgebras corresponding to the superconformal algebras in 
$d\leq 6$, and verify that there are no superconformal algebras  in $d > 6$
. We will proceed to use a method developed by Dobrev and Petkova in 
\rdpresult , \rdpreview ,\rdpdetails ,  (using the theory set up in \rkacmain , )  
to almost completely determine the constraints imposed by unitarity on representations of these algebras. These constraints will determine the required 
inequalities on the scaling dimensions of local operators in a unitary  QFT. 

	In section 5 we examine how our results fit into what is known about
systems with superconformal invariance. We examine some primary operators
in the world volume theory of the $M_2, D_3 , M_5$ brane, and identify the 
representations that these operators label from the list of allowed 
representations generated in previous section. We then go on to review 
the oscillator construction \rgunaydinc\ of the $d=6,n=2$ algebra, and
explicitly construct the short representations of this algebra whose 
existence could  not be conclusively established in section 4.

	In section 6 we present a detailed listing of all our results. The 
reader interested only in our results  may jump directly to section 6.

\newsec{{\bf The Conformal Group.}}

In this section we will discuss the unitary representations of the conformal
group in arbitrary dimension, as relevant to Quantum Field Theory. Our chief
result will be the following. Primary fields in a conformally invariant QFT appear
in multiplets that form representations of $SO(d)$. They also have a scaling
dimension, $\epsilon_0$. 
Unitarity forces $\epsilon_0$ to obey an inequality of the form
\eqn\confineq{ \e \geq f(representation\  \ of \  \ field) }
where $f$ is a function we determine below. 
We will also obtain restrictions on the $SO(d)$ content of free operators in 
a conformal QFT. 

\subsec{{\bf Definition of the Conformal Group }}

The conformal group in $d$ dimensions is generated by ${d(d-1) \over 2}$ Lorentz generators
$M_{\mu\nu}$, $d$ momenta $P_{\mu}$, $d$ special conformal generators $K_{\mu}$ 
and a dilatation $D$. (Through this section greek indices run from 0 to $d-1$).
These quantities obey commutation relations
\eqn\mmcom{ [M_{\mu\nu} , M_{\alpha\beta}] = (-i)[ \eta_{\mu\beta}M_{\nu\alpha} + \eta_{\nu\alpha}M_{\mu\beta} - \eta_{\mu\alpha}M_{\nu\beta} - \eta_{\nu\beta}M_{\mu\alpha} ]}
\eqn\mpcom{ [M_{\mu\nu}, P_\alpha] = (-i)[\eta_{\nu\alpha} P_\mu - \eta_{\mu\alpha}P_{\nu}]} 
\eqn\mdcom{ [D, M_{\mu\nu}] = 0}
\eqn\mkcom{ [M_{\mu\nu}, K_\alpha] = (-i)[\eta_{\nu\alpha} K_\mu - \eta_{\mu\alpha}K_\nu]}
\eqn\dpcom{ [D,P_\mu] = -ik_\mu}
\eqn\dkcom{ [D,K_\mu] = -i(-K_\mu)}
\eqn\ddcom{ [D,D] = 0}
\eqn\ppcom{ [P_\mu,P_\nu] = 0}
\eqn\pkcom{ [P_\mu,K_\nu] = (-i)[2\eta_{\mu\nu}D + 2M_{\mu\nu}]}
\eqn\kkcom{ [K_\mu,K_\nu] = 0}
We are interested in representations of the conformal algebra in which all 
the generators above are implemented as hermitian operators.\foot{ It is 
easy to check that this is compatible with the commutation relations \mpcom\ -
\kkcom\ , 
by checking that the later are invariant under conjugation.}

A representation of these commutation relations may be obtained through 
differential operators
\eqn\mdiff{ M_{\mu\nu} = (-i)[x_{\mu}\partial_{\nu} -x_{\nu}\partial_{\mu}]}
\eqn\pdiff{ P_{\mu} = (-i)[\partial_{\mu}]}
\eqn\kdiff{ K_{\mu} = (-i)[-2x_{\mu}(x.\partial) + x^2\partial_{\mu}]}
\eqn\ddiff{ D = (-i)[-x.\partial]}
The Conformal Group group is locally isomorphic to the group $SO(d,2)$. 
Denote $SO(d,2)$ generators by $S_{ab}$ where
latin indices run from -1 to $d$. The indices -1 and 0 are associated with
-1 in the metric. The isomorphism of $SO(d,2)$ and the standard conformal algebra
is given by 
\eqn\sodm{ S_{\mu\nu} = M_{\mu\nu}}
\eqn\sodd { S_{-1d} = D }
\eqn\sodpk{ S_{\mu-1} = \half [ P_{\mu} + K_{\mu}]}
\eqn\sodkp{ S_{\mu d} = \half [ P_{\mu} - K_{\mu}] }
One may check that $S_{ab}$ thus defined obey commutation relations 
analogous to \mmcom , with $Ms$ replaced by $Ss$. 

It is useful to define auxiliary quantities that obey the commutation rules 
of a Euclidian conformal algebra. Below indices $p$ and $q$ range from 1 to 
$d$.
\eqn\newm{ M'_{pq} = S_{pq}}
\eqn\newd{ D' = (i)S_{-10}}
\eqn\newp{ P'_p = [S_{p-1} + i S_{p0}]}
\eqn\newk{ K'_p = [S_{p-1} -iS_{p0}]}
The commutation relations of these new objects are those of the generators of 
the Euclidian conformal group $SO(d+1,1)$. Primed objects have hermiticity 
properties derived from those of the normal generators
\eqn\mconp{M'^{\dagger} = M'} 
\eqn\dconp{D'^{\dagger} = -D'}  
\eqn\pconp{P'^{\dagger} = K'}
\eqn\kconp{K'^{\dagger} = P'}
Of course the transition from the Lorentzian conformal group to the Euclidian 
conformal group was made with the help of $i$ s, this shows up in the conjugation 
relations \mconp\ - \kconp\ .

\subsec{{\bf Interpretation of Primed operators}}

We will now give an interpretation of the primed operators we have defined 
 in \newm\ - \newk\ , and explain why they, rather than the true generators
of the algebra, will enter most of the calculations in this paper.
 Consider a Conformal QFT in $(d-1,1)$ dimensions.
Consider a Euclidian continuation (Wick rotation ) of this theory. 
Quantize this Wick rotated theory radially, i.e., choose as surfaces
of constant time the $d-1$ dimensional spheres centered around the origin. 
Interpret
the sphere at infinity and at the origin respectively as the infinite future 
and
the infinite past. The Euclidian 
QFT thus constructed has invariance under the conformal group $SO(d+1,1)$, 
whose generators we will call $P'_p,\  \ M'_{pq},\  \ K'_{p}$.\foot{ To forestall
confusion we state explicitly that $P'$ etc., 
are not related to the generators of symmetry of the
Lorentz version of the same theory through \newm\ - \newk\ }
Here, for instance,  $P'_m$ 
is the operator that generates motion in the $m^{th}$ spatial direction 
(one of the $d$ Cartesian directions in our $d$ dimensional space). 

Not all $SO(d+1, 1)$ generators 
are Hermitian in this theory. Because the surfaces of `constant time' are
spheres, $M'$ operators are guaranteed to be Hermitian and so obey \mconp . 
Since surfaces of constant `x' are not equal time surfaces, $P'$ operators are
not necessarily Hermitian; in fact one may show that, (as in $d=2$ CFTs), radial 
quantization imposes the Hermiticity condition \pconp , \kconp , on $P'$ and 
$K'$ operators (c.f. $L_1$ and $L_{-1}$ in d=2 CFT). Similarly the `Hamiltonian'
of our theory, i.e., the generator of scale transformations $D'$, is antihermitian
 (reflecting the fact that we are working in a Euclidianized theory), and so 
obeys \dconp . In net, then, the Euclidian theory radially quantized has a 
symmetry group generated by $SO(d+1,1)$ generators that obey Hermiticity 
conditions \mconp\ - \kconp . 

This is reassuring. The Lorentzian theory is invariant under $SO(d,2)$, 
whereas its Wick rotation has invariance under $SO(d+1,1)$. This may worry the 
reader who is aware
of the big difference between the unitary representations of SO(d,2) and 
$SO(d+1,1)$; merely from group theory it seems that the two theories describe
the same physics. This `paradox' has the obvious resolution ; the Euclidian theory 
does not represent $SO(d+1,1)$ unitarily; generators of the radially quantized theory 
obey \mconp\ - \kconp\ , and $SO(d+1,1)$ with generators obeying these commutation
relations  is identical to $SO(d,2)$ with unitary generators, definitions
 \newm\ -\newk\ and consequent equations \mconp\ - \kconp\ show.

Therefore, to study constraints imposed on the scaling dimensions of operators
in a conformal QFT one may adopt two equally valid procedures. The first and 
most straightforward would be to study the Lorentzian theory - with an 
$SO(d,2)$ symmetry unitarily implemented, and study the restrictions on the 
eigenvalues of operators under scalings D, i.e., on $\e$ defined by [D, Op] = (-i)$\e$ Op. A second, and equally valid procedure - the one we adopt in this 
paper - 
is to study the Euclidianized version of the theory that has $SO(d+1,1)$ symmetry
implemented non-unitarily as in \mconp\ - \kconp\ , and study the restrictions
on the eigenvalues of $D'$ - interpreted as a scaling dimension of the operator
in question only after a Wick rotation. That is we study 
restrictions on scaling dimensions $\e$ of operators
defined by $[D',Op']=(-i)(-i\e Op')$. 

\subsec{{\bf The structure of unitary representations}}

Physically interesting irreducible unitary representations of the Conformal algebra are
modules with energy bounded from below.\foot{ As we are 
interested in the implementation of this symmetry in a quantum theory, we actually 
describe irreducible representations of generators of the covering group. We will use notation loosely through the paper. In particular, when we speak of $SO(d)$ we
always actually refer to its covering group}
The structure of such
modules is described great detail in \revans\ for the special case $d=3$, 
and in \rmack\ for the case $d=4$. Using the arguments in these papers one 
may deduce the following structural features of physically interesting modules
 in arbitrary dimension. 

A unitary irreducible representation of the conformal group may be decomposed into a direct sum
 of irreducible representations of its maximal compact subgroup - $SO(d)\times SO(2)$. The $SO(d)$
here is generated by $M'$, and the $SO(2)$ by $D'$.  No given
$SO(d)\times SO(2)$ representation occurs more than finitely many times in
 this decomposition. The scaling dimension ($D'$ eigenvalue) of states in 
representations of physical interest are all of the form  $\epsilon_0 $ + n, 
where n is a positive integer. $\e$ is the lowest dimension in the module, and 
is referred to as the scaling dimension of the module. The conformal module, if
it is irreducible, contains exactly one $SO(d)\times SO(2)$ irreducible representation with 
dimension $\e$. Let this representation have $SO(d)$ lowest weights  
$(-h_1..-h_{[d/2]})$. These are referred to as the $SO(d)$ lowest  weights of our 
module and the lowest $SO(d)$ weight vector in the $SO(d)$ irreducible representation with dimension 
$\e$ 
is called the lowest weight vector of the conformal module. The full conformal
module is uniquely determined by specifying its scaling dimension and its
$SO(d)$ lowest weights. All states in the conformal module may be generated by
acting on states of scaling dimension $\e$ by $P'$ operators (which raise the scaling 
dimension of states by unity).

\subsec{ {\bf Implications for operators in a Conformal QFT}}

In the previous subsection we described the structure of a unitary representation
of the conformal algebra. It is composed of a module of states, and is 
completely characterized by a lowest weight state. We are interested in 
the application of this theory to the representations of the conformal algebra
on the space of local quantum operators in a conformal QFT. The `states'
in our module will be local quantum operators. The action of conformal 
generators on these `states' will be given by the commutator. 
Lowest weight operators will occur in multiplets that fall in an 
irreducible representation of $SO(d)\times SO(2)$. Lowest weight operators
(or primary operators, as we will sometimes call them) commute with $K'$ 
(lowering) operators. All operators in a given module may be obtained 
by commuting primary operators with $P's$.

We can be more explicit. Consider a multiplet of local operators, 
$\phi_{\alpha}(x)$ in a conformally invariant QFT. The x dependence of these
operators can be solved for; $\phi_{\alpha}(x) = 
e^{-iP.x}\phi_{\alpha}(0)e^{iP.x}$. (Here $P$ is the generator of translations 
in our space. In the Euclidian theory $P$ should be replaced by $P'$). 
Differentiating this expression one obtains 
$[P_{\mu} , \phi_{\a}(x)]=(-i)(-\partial_{\mu}\phi_{\a}(x))$. We may also 
determine the form of the commutator of $\phi_{\alpha}(x)$ with the other
 generators
of the conformal group.\foot { Let $T_a$ stand for the set of generators of 
the conformal algebra.  Let
$[T_a,e^{-iP.x}] = t_a e^{-iP.x} + e^{-iP.x}\b_a^b T_b$ (where $t_a$ is a 
hermitian differential operator and $\b$ a matrix). Then 
$[T_a,\phi_{\a}(x)] = t_a\phi_{\a}(x) + e^{-iP.x}[T_a +\b_a^bT_b , \phi_{\a}(0)] e^{iP.x}$. }
We specify the transformation properties of our multiplet by
setting $[M_{\mu\nu},\phi_{\a}(0)]= (-M_{\mu\nu}^R)_{\a}^{\b}\phi_{\b}(0)$,
and     $[D,\phi_{\a}(0)]= \e\phi_{\a}(0)$, 
where $M^R$ is a the Lorentz Transformation operator in the R representation -
the representation $\phi$ happens to transform in and $\e$ is the scaling 
dimension of $\phi$. The action of conformal generators
on operator $\phi$ is then given by differential/ matrix operators acting from 
the left as
\eqn\mcoset{ M_{\mu\nu} = (i)[x_{\mu}\partial_{\nu} -x_{\nu}\partial_{\mu}]-M^R_{\mu\nu}}
\eqn\pcoset{ P_{\mu} = (i)[\partial_{\mu}]}
\eqn\kcoset{ K_{\mu} = (i)[-2x_{\mu}(x.\partial) + 
x^2\partial_{\mu}]-2x^{\a}M^R_{\a\mu} +2x_{\mu}\e}
\eqn\dcoset{ D = (i)[-x.\partial] + \e}
The difference in signs between \mcoset\ - \kcoset\  and \mdiff\ -\kdiff\  is 
the 
usual coset space artifact (derivatives act from the 'inside' rather than the 
`outside'). A primary operator is killed by the $K$ ($K'$ in the Euclidian theory) differential operator above. 
 
\subsec{ {\bf Restrictions imposed by Unitarity}}
 
In this subsection we will derive restrictions placed by unitarity on the scaling
dimension of an acceptable module as a function of its highest $SO(D)$ weights.

Let $|\{s\}\rangle \  \  (\{s\} = \{ s_1 , s_2 , ... s_n \}$ where $n=[d/2]$, 
$s$'s are $SO(d)$ weights), represent the kets of the $SO(d)$ irreducible representation that 
contains the lowest weight state of our module. All the states above 
have -i$D'= \epsilon_0$. Let the representation hosted by $|\{s\}\rangle$ have 
highest weights {h} =$\{h_1, h_2, ... h_n\}$.

Consider the states $ P'_{\mu}|\{s\}> $. This is the set of 
states in the with dimension $\epsilon_0 +1$. If our module is unitary, 
arbitrary linear combinations of these states must have positive
norm. Using the fact that $P'^{\dagger} = K'$,
the above condition is equivalent to demanding that the matrix
 \eqn\mat{ A_{\mu\{s\} , \nu\{t\}} =<\{s\}|K'_{\mu} P'_{\nu}|\{t\}>}
is positive, i.e., has only non negative eigenvalues. Using the group
commutation relations \pkcom, and $D' = (-i)\epsilon_0$
 \eqn\nmat{ A_{\mu\{s\} , \nu\{t\}} = 2<\{s\}|\epsilon_0 + (-i)M'_{\mu\nu}|\{t\}>}
Positivity of A is then the condition that the matrix
 \eqn\nnmat{B_{\mu\{s\} , \nu\{t\}} = <\{s\}|(-i)M'_{\mu\nu}|\{t\}>}
has no eigenvalue smaller than $-\epsilon_0$. 

To process this condition we use a trick. Notice

\eqn\trick{(-i) M'_{\mu\nu} = \half(-i)(\delta_{\mu\alpha}\delta_{\nu\beta} -
\delta_{\mu\beta}\delta_{\nu\alpha}) M'_{\alpha\beta} }
That is \foot{In the equation below $M$ is an 
operator, $V$ is a matrix, $\mu$ and $\nu$ are its indices.} 
\eqn\trickp{ M'_{\mu\nu} = (V \cdot M')_{\mu\nu}}
Where $(V_{\alpha\beta})_{\mu\nu} = (-i)(\delta_{\mu\alpha}\delta_{\nu\beta} -
\delta_{\mu\beta}\delta_{\nu\alpha}) $ are the $S0(d)$ generators in the vector
representation, and $A \cdot B$ is defined as $\half A_{\alpha\beta}B_{\alpha\beta}$, 
for A and B generators of $SO(d)$ in arbitrary representations, both sides of
this definition being operators (matrices) on the tensor product of the two 
representation spaces. Therefore
\eqn\trickpp{  B_{\mu\{s\},\nu\{t\}} = (V \cdot M')_{\mu\{s\},\nu\{t\}} }
Now
\eqn\trickppp{  V \cdot M' = \half( (V +M')\cdot (V+M') -V \cdot V - M' \cdot M')}
Determining the allowed eigenvalues of this operator is now analogous to 
the problem of determining the allowed values of $L\cdot S$ for the hydrogen atom.
One may perform a similarity transformation on our matrix $B$ that
corresponds to moving to a Clebsch Gordan coupled basis in the tensor product
space. In such a basis each of $ (V+M')^2 , V^2$, and $M'^2 $ are `good quantum
numbers'.

Let the $M'$ matrices transform in representation $R$, and let $R'$ be the 
representation
with smallest quadratic Casimir\foot{$c_2(R') = (defn) M.M$ (in the $R'$ rep)} 
that appears in the Clebsch Gordan expansion of $R\times V$.
Then the condition \nnmat\ is 
\eqn\confcond{\epsilon_0\geq \half(c_2(R) + c_2(V) -c_2(R'))}
To process this equation it is helpful to note the following formula for the
quadratic casimir of $SO(d)$ in terms of its highest weights
\eqn\casimir{c_2(\{h\}) = \sum_{i=1}^{[d/2]} (h_i^2 +(d-2i)h_i)}
as well as the following special cases; $c_2(scalar) = 0$, 
$c_2(spinor) = (d/8)(d-1)$, ${c_2(vector) = d-1}$.
The highest weights of $R'$ in terms of those of $R$ may be deduced from appendix
2 and the condition above made more explicit. For instance, if the 
representation R has highest wts s.t. $h_1 \geq |h_2|+1$
then the condition above becomes ${\epsilon_0\geq h_1 + d-2}.$
The formula for an arbitrary representation in terms of its highest weights is
\eqn\confcondc{\epsilon_0\geq |h_i| + d-i-1} 
where $i$ is the smallest value s.t. $h_i \geq |h_{i+1}|+1$, provided there is
such an $i$. If no such $i$ exists then all lowest weights have equal modulus value
. In the case that they are all equal to zero the condition is merely 
$\e \geq 0$. If they are all equal and $\geq$ 1 then the condition above 
applies with $i=[d/2]$. The only remaining case is that of spinor representations
for which the answer is $\e \geq (d-1)/2$. The 
answer for the vector representation is $\e \geq d-1$. We make some special cases explicit below

In $d = 3$ representations of $SO(d)$ are labeled by a half integer $j$. 
The condition
above is simply
\eqn\confcondt{\epsilon_0 \geq 0 \  \ (j=0)}  
\eqn\confcondtt{\epsilon_0 \geq 1 \  \ (j=1/2)} 
\eqn\confcondttt{\epsilon_0 \geq j+1 \  \ (j\geq 1)}
These are exactly\foot{These results supplemented with the level 2 condition
ahead for the j=0 case} 
the conditions derived in \revans . 
In $d= 4$ $SO(4) = SU(2)\times SU(2)$, and so representations are labeled by two 
half integers,  $j_1$ and $j_2$. The conditions derived for this case is 
\eqn\confcondf{\epsilon_0 \geq f(j_1) + f(j_2) } 
Where $f(j)$ is defined by
\eqn\denf{ f(j)=0\  \  for\  \  j=0,\  \ f(j) = j+1 \  \ for \  \ j>0}
These are precisely\foot{These results supplemented with the level 2 condition
ahead for the j=0 case}  the conditions derived in \rmack\ .

In $d=5$ representations are labeled by $SO(5)$ highest weights $h_1$ and $h_2$. 
The conditions are 
\eqn\confcondf{\epsilon_0 \geq 0 \  \ (h_1=h_2=0)}
\eqn\confcondfs{\epsilon_0 \geq 2 \  \ (h_1=h_2=\half)}  
\eqn\confcondff{\epsilon_0 \geq h+2 \  \ (h_1=h_2=h\neq 0,\half)} 
\eqn\confcondfff{\epsilon_0 \geq h_1+3 \  \ (h_1>h_2)}
In $d=6$ representations are labeled by $SO(6)$ highest weights $h_1,h_2$ and $h_3$. The conditions are
\eqn\confconds{\epsilon_0 \geq 0 \  \ (h_1=h_2=h_3=0)}
\eqn\confcondss{\epsilon_0 \geq h+2 \  \ (h_1=h_2=|h_3|=h \neq 0)}
\eqn\confcondssss{\epsilon_0 \geq h+3 \  \ (h_1=h_2 >|h_3|)} 
\eqn\confcondsssss{\epsilon_0 \geq h_1+4 \  \ (h_1>h_2)}
In an arbitrary dimension $ d$ special representations obey
\eqn\confcondgs{\epsilon_0 \geq 0 \  \ (scalar) }
\eqn\confcondgsp{\epsilon_0 \geq (d-1)/2 \  \ (spinor)}
\eqn\confcondgv{\epsilon_0 \geq (d-1) \  \ (vector)}

All these results make eminent sense. Since the identity operator
in our field theory constitutes a singleton scalar representation with 
$\epsilon_0 =0$ we could not have expected a stronger bound for the scalar 
case. \foot{Note that any operator that has $\e$=0 is annihilated by (i.e., commutes with)
the momentum operator $P'_{\mu}$, and so is translationally invariant. The
identity operator is the only translationally invariant local operator, and so
it constitutes the unique representation with $\e =0$}

The operator that saturates the bound for the spinor representation  
gives zero on contracting with  $P'_{\mu}$ in 
such a manner as to get the smallest resultant representation. 
Therefore this operator obeys $[\pslash ', \psi] =0$, i.e., the Dirac equation. 
Therefore our bound is saturated by the Free Dirac field, and 
indeed the scaling dimension above is the canonical dimension for 
such a field.   

The bound on the vector representation is, at first sight, a little puzzling.  
In 4 dimensions, for instance, it is $\epsilon_0 \geq 3$, this seems to be
in contradiction with the fact that the scaling dimension of the vector field
in Maxwell theory is 1. The contradiction is explained away by noting that the
Maxwell $A_{\mu}$ field is not gauge invariant, and so is not represented in 
a theory with only positive norm states. Our restrictions apply only to
gauge invariant operators. Vector operators that saturate the bound above 
satisfy $[P_{\mu},\psi_{\mu}] = 0$; 
examples of such operators are conserved currents; these are indeed vectors,
 and indeed have the scaling dimension above.

We have obtained constraints from a unitarity analysis at level one. New 
constraints may well emerge at level\foot{The level of $|\psi \rangle$ is the 
number of $P'$ operators that act on the lowest weight  state to give 
$|\psi \rangle$. When we deal with the superconformal algebra, `level' will 
have the same meaning but with $P'$ replaced by $Q'$.}  2 and higher.
The level 2 analysis
is particularly easy to carry out for a lowest weight state in the scalar 
representation and yields 
 \eqn\scalarfin{\epsilon_0(\epsilon_0 -(d-2)/2)  \geq 0 \  \ (scalar).}
This condition permits the singleton representation with $\e =0$, but forces 
all non singleton scalar representations to have dimension greater than the 
scaling dimension of the free scalar field.

We now have a set of necessary conditions on scaling dimensions of primary 
operators in a unitary conformal quantum field theory. Nothing we have said
so far has given us any reason to suspect that this is also a sufficient set
of conditions. However comparison with the extensive analysis carried out
in \revans\ and \rmack\ show that the conditions we have turn out to
be both necessary and sufficient to ensure unitarity in $d=3,4$. 
This may or may not be a 
coincidence of low dimensions. It would be interesting - and seemingly not 
too difficult - to attempt to generalize
Mack's construction of explicitly unitary representations in \rmack\ to higher
dimensions, and so obtain sufficient conditions for unitarity.

\subsec{ {\bf Free Conformally invariant fields}}

One often thinks of a conformally invariant theory as a theory of massless
particles. Since we know that $free$ massless particles in d dimensions appear
in $SO(d-2)$ multiplets, it is perhaps a little puzzling that operators in a
conformally invariant theory appear in multiplets of $SO(d)$.  

To understand this start with \mcoset\ -\dcoset\ for the action of 
conformal generators on an arbitrary $SO(d)$ multiplet of operators. 
Impose the condition $P.P \phi_{\a}(x)=0$ on all operators in the multiplet. 
Since this equations is true for all $\phi$s in the multiplet, making an
infinitesimal conformal transformation on any specific $\phi$ gives
$P.P T_a\phi_{\a}(x)=0$ for all conformal generators $T_a$.  Which means that
\eqn\fcond{[P.P,T_a] \phi_{\a}(x)=0}
 for all conformal generators $T_a$. Note that P.P is not a casimir of the 
conformal group. 

Setting $T_a=P_{\mu}, M_{\mu\nu} , D$ in \fcond\ yields trivialities.   
Setting $T_a=K_{\mu}$ and using \mcoset\ - \dcoset\ yields the following 
restriction on $\phi_{\a}(p)$ (we have fourier transformed the operator). 
\eqn\free{ p_{\mu}(-iM^R_{\mu\nu})_{\a}^{\b}\phi_{\b} = ({d-2\over 2}-\e) p_{\nu}
\phi_{\a} } To analyze this equation get into a frame in which 
$p_{\mu}=(1,1,0,0,0..0)$, where the components in (   ) are time, the first
spatial component,  ... . \free\  is $d$ equations because of the free
index $\nu$. When $\nu$ = 2, 3 ..d-1 , the RHS of \free\  is zero, and 
the equation becomes
\eqn\freee{(M^{R}_{1i}-M^{R}_{0i})\phi =0}
$Ms$ here are matrices representing $SO(d-1,1)$. Convert these to 
$SO(d)$
generators by $M^{lorentz}_{\mu\nu}=(-i)^{n} M^{Euclid}_{\mu\nu}$, where n is 
the number of zeroes in $\mu,\nu$.  
Let the first weight of our $SO(d)$ algebra be that associated with $M^R_{01}$,
\freee\ says that any $M^R$ that raises the first weight of $\phi$ and 
does absolutely anything to the other weights kills $\phi$. That is, 
$\phi$ is a state of highest first weight in the $SO(d)$ multiplet, but whose 2nd to $[d/2]^{th}$ weights can take any value in the multiplet. 
When we set $\nu$ = 0 or 1 \free\  becomes 
$(-i)M^R_{01}\phi =({d-2\over 2}-\e)\phi$. Since
$M_{0i}^{lorentz}=(-i)M_{0i}^{Euclid}=H_1$ ($H_1$ is the
first Cartan generator of $SO(d)$), we obtain
\eqn\freet{\e=h_1+{d-2\over 2}}
where $h_1$  is the highest first weight of $\phi$. The analysis above 
has given us an 
equation for the scaling dimension of $\phi$ as a function of its $SO(d)$ 
representation. It has also made clear the connection  between the $SO(d)$ 
representation of the conformal operator, and the $SO(d-2)$ little group 
representation of the particle it creates. Explicitly, for a free conformal
operator in momentum space, the $SO(d)$ weight
connected with rotations in the $momentum - time$ plane is restricted to be
the highest weight in the $SO(d)$ representation. 
The remaining weights of $\phi$ are arbitrary, and in fact fill out a multiplet
of $SO(d-2)$ - the little group of the massless Lorentz algebra. Therefore
a free conformally invariant
field (if one such exists-see below) with $SO(d)$ highest weights 
$h_1, h_2 ..h_{[d/2]}$
 corresponds to a particle transforming in $h_2 , h_3 , ... h_{[d/2]}$ of its 
$SO(d-2)$ little group. For instance, an $SO(d)$ scalar is 
an $SO(d-2)$ scalar, an $SO(d)$ spinor of given chirality is an $SO(d-2)$ spinor
of the same chirality, an $SO(d)$ vector is an $SO(d-2)$ scalar, and an $SO(d)$
antisymmetric tensor is an $SO(d-2)$ vector. (We will actually see that free 
representations corresponding to the last 2 do not exist.)

We have noted above that a multiplet $\phi_{\a}$ that obeys
$P.P \phi_{\a}(x)=0$ has also, for consistency, to
obey $[P.P,T_a] \phi_{\a}(x)=0$. That is not the only consistency condition,
however. One may iterate the argument above to generate the conditions
\eqn\scond{[[P.P,T_a],T_b] \phi_{\a}(x)=0}
 for arbitrary conformal generators $T_a$, 
$T_b$, and so on. These equations yield new equations for $\e$ which may or 
may not be consistent with \freet . An inconsistency would rule out the 
existence of a free operator in a given representation. Choosing $T_a, 
T_b$ above to be $K_{\a} , K_{\b} $ and using \mcoset\ -\dcoset\  we obtain\foot{ This equation was obtained in \rsiegel\ by different means}
\eqn\freef{(-M^R_{\b\mu}M^R_{\mu\a}-M^R_{\a\mu} M^R_{\mu\b}) \phi =
\eta_{\a\b}(2\e)(\e -{d-2 \over 2})\phi} 
 \freef\ may
be processed in detail as an eigenvalue equation, but for our purposes it 
is sufficient to trace it over $\a$ and $b$ indices. This yields 
\eqn\freef{ \e (\e-{d-2\over 2}) -{2\over d} c_{2}(R) = 0}
an equation that could also have been obtained by employing 
the techniques of the previous section by setting 
$<\{s\}|(K' \cdot K)('P' \cdot P')|\{t\}>=0$ .
We may now ask under what circumstances \freet\  and \freef\  are consistent. 
It is easy to check that this is the case only when $h_1=h_2=h_3=...=h_{[d/2]}=h$
for any $h$ (when d is even), and when  $h_1=h_2=h_3= ..=|h_{[d/2]}|=h$ for 
$h=0$ or $\half$ 
(when d is odd). These constraints on possible free representations were obtained
in \rsiegel  .

\newsec{ The Super Conformal Algebras}

The structure of the supersymmetry algebra in arbitrary dimension is well 
known \rstrathdee . While minor details of the algebra vary with dimension, 
the supersymmetry generators $Q_{\alpha}$ always transform as spinors of the 
Lorentz Group. 
In a conformally invariant theory the $(d-1,1)$ Lorentzian Spinor $Q_{\alpha}$
forms a part
of (is completed to) a $(d,2)$ conformal spinor. 

Notice that $(d-1, 1)$ and $(d, 2)$ spinors have 
identical reality properties. In fact a $(d,2)$ conformal spinor with specific 
reality properties is composed of two $(d-1, 1)$ Lorentzian spinors of the 
same 
reality properties.
This can be made explicit by the following choice of $\Gamma$ matrices for
$SO(d,2)$.\foot{ $SO(d-1,1)$ $\Gamma$ matrices are denoted by $\sigma $ through the 
rest of this paper.}
$$\Gamma_{\mu}=\left(\matrix{\sigma_{\mu}&0. \cr
                             0&-\sigma_{\mu}.\cr}\right)$$

$$\Gamma_{-1}=\left(\matrix{0&-I. \cr
                            I&0. \cr}\right)$$

$$\Gamma_{d}=\left(\matrix{0&I. \cr
                            I&0. \cr}\right)$$

$$\Gamma_{d+1}=\left(\matrix{\sigma_{d+1}&0. \cr
                            0&-\sigma_{d+1}. \cr}\right)$$
The last matrix exists only for even $d$.
$Q$ is completed to give a full conformal spinor V with the introduction of a 
new Lorentz spinor $S$ as  
$$V=\left(\matrix{Q_\alpha. \cr
                            C_{\theta \phi}\bar{S}^{\phi} . \cr}\right)$$
C is the charge conjugation matrix.
We set\foot{Recall that if $[M^i,Q_\alpha] = G^i_{\alpha\beta}Q_{\beta}$ then the 
commutation relations between the operators $M^i$ are the same as those between
the matrices $G^{iT}$ (not $G^i$), i.e., those between $-G^i$. We have thus 
chosen G to be  -(matrices that have standard M commutation relations).}
\eqn\mocom{[S_{ab}, V_{\alpha}] = R(M_{ab})_{\alpha}^{\beta} V_{\beta}}
With $R(M_{ab}) =(i/4)[\Gamma_a, \Gamma_b]$.
Specifically
$$R(P_{\mu})= (-i)\left(\matrix{0&0 \cr
                            \sigma_{\mu}&0 \cr}\right)$$
$$R(K_{\mu})= (-i)\left(\matrix{0& \sigma_{\mu} \cr
                            0&0 \cr}\right)$$
$$R(D)= (-i/2)\left(\matrix{I&0 \cr
                            0&-I \cr}\right)$$
$$R(M_{\mu\nu})= \left(\matrix{R(m_{\mu\nu})&0 \cr
                            0&R(m_{\mu\nu}) \cr}\right)$$
Where $R(m_{\mu\nu}) = (+i/4)[\sigma_{\mu}, \sigma_{\nu}]$\foot
{Notice this means that $Q$ and $C\bar{S}$
each transform as spinors under the Lorentz group, have dilatation wt $\half$
 and $-\half$ respectively, and, schematically,  $[P,Q] =[K,C\bar{S}] = 0$, 
$[K,Q]=C\bar{S}$, $[P,C\bar{S}] =Q$.}

	When working with the conformal group we had found it useful to define
auxiliary (primed, non hermitian) generators \newm\ - \newk\  that obeyed
the commutation relations of generators of the Euclidian conformal group. 
We interpreted these operators physically in subsection 2.2. We will now 
define objects that will be similarly useful, and that have  an equivalent
interpretation in the full superconformal group;
 auxiliary odd generators $Q'$ and $S'$, objects
that transform as Euclidian spinors under our auxiliary (primed) conformal
generators. Define\foot{The fact $Q$ is a chiral spinor makes the definition
of $Q'$ and $S'$ different in d=6. In that dimension as in all others 
$Q'$ and $S'$ are defined so as to make
$M'$ and $D'$ block diagonal in the $Q'$, $S'$ basis.} 
\eqn\qprime{ Q' = {1 \over \sqrt{2} }(Q - i\sigma_0 S)}
\eqn\qprimee{ S' = {1 \over \sqrt{2} }(Q + i\sigma_0 S)}

The action of $M' , P' , K'$ etc., on $Q'$ and $S'$ may now be determined.

\eqn\mqcom{ [M'_{pq},Q'_{\alpha}] = (i/4)[\Gamma_p, \Gamma_q]_\alpha ^\beta 
Q'_{\beta}}
\eqn\mscom{ [M'_{pq},S'_{\at}] = (i/4)[\Gt_p, \Gt_q]_{\at}^{\bt} 
Q'_{\bt}}
\eqn\dqcom{ [D',Q'_{\alpha}] = (-i/2)Q'_{\alpha} }
\eqn\dscom{ [D',S'_{\at}] = (-i/2)-S'_{\at} }
\eqn\dqcom{ [P'_p,Q'_{\alpha}] = 0}
\eqn\dqcom{ [K'_p,S'_{\at}] =0}
\eqn\pscom{ [P'_p, S'_{\at}] = -(\Gt_p \sigma_0)_{\at}^{\beta}Q_{\beta}} 
\eqn\pscom{ [K'_p, Q'_{\alpha}] = +(\Gamma_p \sigma_0)_{\alpha}^{\bt}S_{\bt}} 
Where\foot{$\Gamma$ and $\Gt$ matrices each generate a Euclidian Clifford 
algebra in $d$
dimensions. $Q'$ transforms as a spinor in the $\Gamma$ representation,
and $S'$ as a spinor in the $\tilde{\Gamma}$ representation.
Notice that in a dimension  in which Lorentzian $\sigma $ matrices may all be 
chosen real, (and making such a choice), the two sets of $\Gamma$ matrices 
above are complex conjugates of each other, and so $Q'$ and $S'$ transform in 
complex conjugate representations. Notice also that $\sigma_0$ interpolates 
between ordinary and tilde indices, as is clear from its commutation 
properties with $\Gamma$ and $\Gt$.}
\eqn\gam{ \Gamma_i = \sigma_i \  \ \Gamma_d = -i\sigma_0}
\eqn\gamaa{ \tilde{\Gamma_i} = \sigma_i \  \ \tilde{\Gamma_d} = +i\sigma_0}

To complete the superconformal algebra, we now need to specify anticommutation
relations between various $Qs$ and $Ss$. In general we deal with extended 
susy and so should also add R symmetry indices to $Qs$ and $Ss$, and 
specify their commutation relations with generators of R symmetry also.

Using well known  $Q-Q$  and $R-Q$ (anti)commutation relations \rstrathdee , it is  
straightforward to deduce the commutation relations of $S$ with conformal, 
R symmetry and $Q$ generators, using Jacobi identities. Rather surprisingly,
it turns out not to be possible to satisfy these Jacobi 
identities in arbitrary dimensions. Satisfaction
of the identities requires certain $\Gamma$ matrix identities,which are true
only in low dimensions. Superconformal algebras exist only in $d \leq 6$. 

We will explain this fact from the theory of Lie Super Algebras presented
 in the next
subsection. In the rest of this subsection, we complete our listing of the
commutation relations for the superalgebras that do exist.

			\cl {{\bf d=3}  }

$SO(2,1)$ admits a set of real $\sigma$ matrices. We work in this basis. As in 
\rstrathdee, our $\sigma$ matrices are also chosen to be  hermitian, 
except $\sigma_0$ which is 
antihermitian. The charge conjugation matrix is defined by the equation 
$C^-1\sigma C = -\sigma^T$. With our choice of $\sigma_{\mu}$, $\sigma_0$ 
solves this equation, and so may be
identified with C. 
The 
spinors $Q$ and $S$ are both real; in our special basis this means
\eqn\Qdag{ Q^{^{\dagger}} = Q \  \  ; S^{\dagger} = S}
\eqn\Qdag{ Q'^{\dagger} = S' \  \  ; S'^{\dagger} = Q'}
The algebra exists with an arbitrary number n of susy generators, which are 
distinguished by an R symmetry label $i$. We sometimes suppress indices of 
$Q_{i\alpha}$ depending on context. The R symmetry group is $SO(n)$;
its generators are antisymmetric tensors $I_{ij}$ that obey
\eqn\IIcom{ [I_{ij} , I_{mn}] = (-i)[ I_{in}\delta_{jm}+ I_{jm}\delta_{ij} -
I_{im}\delta_{jn} -I_{jn}\delta_{im}]}
\eqn\IQcom{ [I_{ij} , Q_{m}] = (-i)[ Q_{i}\delta_{jm}+ -Q_{j}\delta_{im}]}
\eqn\IQncom{ [I_{ij} , Q'_{m}] = (-i)[ Q'_{i}\delta_{jm}+ -Q'_{j}
\delta_{im}]}
\eqn\IScom{ [I_{ij} , S_{m}] = (-i)[ S_{i}\delta_{jm}+ -S_{j}\delta_{im}]}
\eqn\ISncom{ [I_{ij} , S'_{m}] = (-i)[ S'_{i}\delta_{jm}+ -S'_{j}
\delta_{im}]}
\eqn\IMcom{ [I_{ij} , M_{pq}] = 0}
Odd elements anticommute according to 
\eqn\qqcom{ \{ Q_{i\alpha}, Q_{j\beta}\} =(\pslash  C)_{\alpha\beta}
 \delta_{ij}}
\eqn\sscom{ \{ S_{i\alpha}, S_{j\beta}\} =(\kslash C)_{\alpha\beta} \delta_{ij}}
\eqn\sqcom{ \{ Q_{i\alpha}, S_{j\beta}\} ={\delta_{ij} \over 2}[(M_{\mu\nu}\Gamma_{\mu}\Gamma_{\nu} C)_{\alpha\beta}   + 2DC_{\alpha\beta}] -
C_{\alpha\beta}I_{ij}}
Primed odd variables obey similar equations 
\eqn\qqncom{ \{ Q'_{i\alpha}, Q'_{j\beta}\} =(\pslash ' C)_{\alpha\beta} \delta_{ij}}
\eqn\ssncom{ \{ S'_{i\at}, S'_{j\bt}\} =(\tilde{\kslash}' C)_{\at\bt} \delta_{ij}}
\eqn\qsncom{ \{ Q'_{i\alpha}, S'_{j\bt}\} =i{\delta_{ij} \over 2}[(M'_{\mu\nu}\Gamma_{\mu}\Gamma_{\nu} C)_{\alpha\bt} + 2D'\delta_{\alpha\bt}]  
-(i)\delta_{\alpha\bt}I_{ij}}

			\cl{  {\bf d=4}  }

d=4 is very similar to d=3. Our $\sigma$ 
matrices are chosen to be real and hermitian (antihermitian for 
$\sigma_0$). Charge conjugation is defined as for d=3, and once again
$\sigma_0$ fits the bill. 
In our Majorana basis 4 component Majorana spinors $Q$ and $S$ are both real; 
\eqn\Qdag{ Q^{\dagger} = Q \  \  ; S^{\dagger} = S}
\eqn\Qdag{ Q'^{\dagger} = S' \  \  ; S'^{\dagger} = Q'}
The R symmetry in d=4 is $U(n)$ (n = the number of Qs)
which acts differently on the positive and negative chirality parts of $Q$.
The algebra exists for arbitrary n.

We define $P_{\pm} =(I \pm \sigma_5)/2)$.
Note that $(P_+)^T = P_-$;  $P_+^* = P_-$; $(P_+)^{\dagger}=P_+$ .
Commutation relations between $U(n)$ generators and the other generators of 
our algebra are\foot{see appendix 1 for notation}.   
\eqn\TQcom{ [T_{ij} , Q_{m}] = [ P_+Q_{i}\delta_{jm} -P_-Q_{j}\delta_{im}]}
\eqn\TQncom{ [T_{ij} , Q'_{m}] = [ P_+Q'_{i}\delta_{jm} -P_-Q'_{j}
\delta_{im}]}
\eqn\TScom{ [T_{ij} , S_{m}] = [ P_{-}S_{i}\delta_{jm} -P_{+}S_{j}\delta_{im}]}
\eqn\TSncom{ [I_{ij} , S'_{m}] = [ P_{+}S'_{i}\delta_{jm} -P_{-}S'_{j}
\delta_{im}]}
\eqn\TMcom{ [I_{ij} , M_{pq}] = 0}
Odd elements obey the following anticommutation relations.
\eqn\qqfcom{ \{ Q_{i\alpha}, Q_{j\beta}\} =(\pslash C)_{\alpha\beta}
 \delta_{ij}}
\eqn\ssfcom{ \{ S_{i\alpha}, S_{j\beta}\} =(\kslash C)_{\alpha\beta} 
\delta_{ij} }
\eqn\sqfcom{ \eqalign{\{ Q_{i\alpha}, S_{j\beta} \}& ={\delta_{ij} \over 2}
[(M_{\mu\nu}\Gamma_{\mu}\Gamma_{\nu} C)_{\alpha\beta} + 2C_{\alpha\beta}D] 
\cr 
&+ 2(i)(P_{+}C)_{\alpha\beta}T_{ij} \  \ -2(i)(P_{-}C)_{\alpha\beta}T_{ji} \  \ +(i/2)(C\sigma_5)_{\alpha\beta}R  \cr}}

Note that in our current choice of basis, $Q'$ and $S'$ transform in inverse 
representations, that is, a lower index with a tilde is equivalent to a 
raised index with no tilde. This comment should make the index structure of the equation below clearer.
\eqn\qqfncom{ \{ Q'_{i\alpha}, Q'_{j\beta}\} =(\pslash ' C)_{\alpha\beta}
 \delta_{ij}}
\eqn\ssfcom{\{ S'_{i\at}, S_{j\bt}\} =(\tilde{\kslash}'C)_{\at\bt} \delta_{ij}}
\eqn\sqfncom{\eqalign{ \{ Q'_{i\alpha}, S'_{j\bt}\} &=((i)\delta_{ij}/2)[(M'_{\mu\nu}\Gamma_{\mu}\Gamma_{\nu})_{\alpha\bt} + 2\delta_{\alpha\bt}D'] \cr &-
2(P_{+})_{\alpha\bt}T_{ij} \  \ +2(P_{-})_{\alpha\bt}T_{ji} \  \ +\half(\sigma_5)_{\alpha\bt}R \cr}}

			\cl{{\bf d=5 } }

Two new things happen in five dimensions. Firstly, $SO(4,1)$ does not admit
real spinors, and so our $Q$ and $S$ spinors are chosen to be pseudo real, that
is they occur in pairs, and obey
\eqn\pseudorealit{ Q_{i\alpha} = \Omega_{ij}(C\sigma_{0}^T)_{\alpha}^
\beta Q_{j\beta}^{\dagger} }
\eqn\pseudorealitys{ S_{i\alpha} = \Omega_{ij}{(C\sigma_{0}^T)}_{\alpha}^
\beta S_{j\beta}^{\dagger} }
Here $\Omega$ is the  $2n\times 2n$  pseudoreality matrix 
consisting of n diagonal 2 $\times $ 2 blocks, each
of which is $-i\sigma_2$ .
C is the charge conjugation matrix $C\Gamma^T C^{-1} = \Gamma$, obeying
$C^*= -C^{-1}$ , $C = - C^T$.
This implies the following conjugation relation for primed quantities defined
in \qprime, \qprimee . 
\eqn\pseudorealityqn{  Q'_{i\alpha } = 
\Omega_{ij}(C\sigma_{0}^T)_{\alpha}^{\bt} S'^{\dagger}_{j\bt} }
\eqn\pseudorealitysn{  S'_{i\at} = 
\Omega_{ij}(C\sigma_{0}^T)_{\at}^{\beta} Q'^{\dagger}_{j\beta} }
The second new thing that happens in $d=5$ is that the superconformal algebra 
exists only for a single pair of $Qs$ (and $Ss$). 
 This is something we will see a reason for in a section 4, but it may also be 
regarded as a 
fact that will be discovered by anyone who tries to write down an n $\geq$ 2 
algebra that obeys all Jacobi identities.

The R symmetry for this supersymmetry algebra is $Sp(1)=SU(2)$.
Denote R symmetry generators $T_a$, a=1,...,3. They generate the standard 
$SU(2)$ algebra. $Q$s and $S$s transform as spinors under this $SU(2)$.
\eqn\mmm{ [T_a , Q_i] = (-\sigma_a' /2)_i^j Q_j }
\eqn\mmn{ [T_a , S_i] = (+\sigma_a' /2)_i^j S_j }
\eqn\mmm{ [T_a , Q'_i] = (-\sigma_a' /2)_i^j Q'_j }
\eqn\mmn{ [T_a , S'_i] = (-\sigma_a' /2)_i^j S'_j } 
$\sigma$'s above are ordinary Pauli matrices,  the primes have been put to 
distinguish them from $SO(d-1,1)$ $\Gamma$ matrices.\foot{The minuses in 
the above equation have been motivated by considerations
similar to those that forced us to choose $(+i/4)[\Gamma_\mu, \Gamma_\nu]$ 
for the Lorentz transformations of the spinors. See the footnote above 
\mocom\ } 

Odd elements anticommute as
\eqn\qqficom{ \{ Q_{i\alpha}, Q_{j\beta}\} =(\pslash C)_{\alpha\beta}
 \epsilon_{ij}}
\eqn\ssficom{ \{ S_{i\alpha}, S_{j\beta}\} =(\kslash C)_{\alpha\beta} 
\epsilon_{ij} }
\eqn\sqficom{ \{ Q_{i\alpha}, S_{j\beta} \} ={\delta_{ij} \over 2}[(M_{\mu\nu}
\Gamma_{\mu}\Gamma_{\nu} C)_{\alpha\beta} + 2C_{\alpha\beta}D] \  \ -6(i)
(T_a\sigma_a/2)_{ij}C_{\alpha\beta} }
Where we have used  the fact that n=1 to replace
the $\Omega$ matrix by $\epsilon=(i)\sigma_2'$, the antisymmetric 2 $\times$ 2
 matrix.

Primed variables obey
\eqn\qqfincom{ \{ Q'_{i\alpha}, Q'_{j\beta}\} =(\pslash ' C)_{\alpha\beta}
 \epsilon_{ij}}
\eqn\ssfincom{ \{ S'_{i\at}, S_{j\bt}\} =(\tilde{\kslash}' C)_{\at\bt} 
\epsilon_{ij} }
\eqn\sqfincom{ \{ Q'_{i\alpha}, S'_{j\bt} \} =[(\delta_{ij}(i)/2)
[(M'_{\mu\nu}
\Gamma_{\mu}\Gamma_{\nu})_{\alpha}^{\theta} + 2\delta_{\alpha}^{\theta}D'] \  \ +6(T_a\sigma'_a/2)_{ij} \delta_{\alpha}^{\theta}][ (-\epsilon_{kj}\sigma_0C)_
{\theta\bt}] }

The last equation implies that
\eqn\sqfinncom{ \{ Q'_{i\alpha}, {{Q'}_{j}^{\theta}}^{\dagger} \} =(\delta_{ij}(i)/2)
[(M_{\mu\nu}
\Gamma_{\mu}\Gamma_{\nu})_{\alpha}^{\theta} + 2\delta_{\alpha}^{\theta}D] \  \ +6(T_a\sigma_a/2)_{ij} \delta_{\alpha}^{\theta}}

			\cl{ {\bf d=6}}

The algebra in $d=6$ is very similar to that in $d=5$ in structure. Once 
again spinors are pseudo real, that is, obey
\eqn\pseudorealityq{ Q_{i\alpha} = \Omega_{ij}(C\sigma_{0}^T)_{\alpha}^\beta 
Q_{j\beta}^{\dagger} }
\eqn\pseudorealitys{ S_{i\alpha} = \Omega_{ij}(C\sigma_{0}^T)_{\alpha}^\beta 
S_{j\beta}^{\dagger} }
$\Omega$ is defined at the beginning of the section on $d=5$. 
In 6 dimensions we have a choice of 2 inequivalent C matrices to work with.
We choose to work with $C\Gamma^T C^{-1} = -\Gamma$, $C^T = C$, 
$C^* = C^{-1}$.

Primed quantities enjoy the following reality properties.
\eqn\pseudorealityqn{  Q'_{i\alpha} = \Omega_{ij}(C\sigma_{0}^T)_
{\alpha}^{\bt} S'^{\dagger}_{j\bt} }
\eqn\pseudorealitysn{  S'_{i\at} = \Omega_{ij}(C\sigma_{0}^T)_
{\at}^{\beta} Q'^{\dagger}_{j\beta} }
It turns out (and we will explain why in the next section) that, although
in $d=6$ supercharges of different chiralities are inequivalent (because
chiral spinors are pseudoreal not complex, unlike $d=4$) superconformal
algebras exist only when all $Qs$ have the same chirality. We will choose this
to be the positive chirality. Define $P_+ = (1+\sigma_7)/2$. 
Note $P_+^{\dagger}=P_{+} ; C^{-1}P_{+}C = P_{-}$.
Superconformal algebras exist with any (even) number - 2n - of 
$Q$s (paired by $\Omega$ as usual). The R symmetry group is $Sp(n)$. Writing out
the details of the theory for general n would involve unfamiliar looking
formulae in terms of $Sp(n)$ generators. Happily, the cases relevant to 
physics - n=1 and n=2 -can be written more simply as $Sp(1) = SU(2)$, and 
$Sp(2)=SO(5)$. We will list the algebra only for these special cases.

\cl{ {\bf n=1}}

The R symmetry group is $SU(2)$; supercharges transform as spinors under 
this group. 
\eqn\mmmm{ [T_a , Q_i] = (-\sigma_a' /2)_i^j Q_j }
\eqn\mmmn{ [T_a , S_i] = (+\sigma_a' /2)_i^j S_j }
\eqn\mmmp{ [T_a , Q'_i] = (-\sigma_a' /2)_i^j Q'_j }
\eqn\mmmq{ [T_a , S'_i] = (-\sigma_a' /2)_i^j S'_j } 
Odd elements anticommute thus 
\eqn\qqscom{ \{ Q_{i\alpha}, Q_{j\beta}\} =(P_+\pslash C)_{\alpha\beta}
 \epsilon_{ij}}
\eqn\ssscom{ \{ S_{i\alpha}, S_{j\beta}\} =(P_-\kslash C)_{\alpha\beta} 
\epsilon_{ij} }
\eqn\sqscom{ \{ Q_{i\alpha}, S_{j\beta} \} ={\delta_{ij} \over 2}[(M_{\mu\nu}
P_+\Gamma_{\mu}\Gamma_{\nu} C)_{\alpha\beta} + 2(P_+C)_{\alpha\beta}D] \  \ -8(i)
(T_a\sigma_a/2)_{ij} (P_+C)_{\alpha\beta} }
Where $\epsilon$ is the $\Omega$ matrix for n=1, defined as for d=5.
Primed variables obey
\eqn\qqfincom{ \{ Q'_{i\alpha}, Q'_{j\beta}\} =(P_{+}\pslash ' C)_{\alpha\beta}
 \epsilon_{ij}}
\eqn\sssncom{ \{ S'_{i\at}, S'_{j\bt}\} =(P_{+}\tilde{\kslash}' C)_{\at\bt} 
\epsilon_{ij} }
\eqn\sqsncom{ \{ Q'_{i\alpha}, S'_{j\bt} \} =[(\delta_{ij}(i)/2)
[(M'_{\mu\nu}
P_{+}\Gamma_{\mu}\Gamma_{\nu})_{\alpha}^{\theta} + 2(P_{+})_{\alpha}^{\theta}D'] \  \ +8(T_a\sigma'_a/2)_{ij} (P_{+})^{\theta}_{\alpha}]
[ (-\epsilon_{kj}\sigma_0C)_{\theta\bt}] }

The last equation implies that
\eqn\sqsnncom{ \{ Q'_{i\alpha}, Q'^{\theta \dagger}_j \} =(\delta_{ij}(i)/2)
[(M'_{\mu\nu}
P_{+}\Gamma_{\mu}\Gamma_{\nu})_{\alpha}^{\theta} + 2(P_{+})_{\alpha}^{\theta}D'] \  \ +8(T_a\sigma'_a/2)_i^j (P_{+})^{\theta}_{\alpha}]}
				
\cl{ {\bf n=2} }

The algebra for the n=2 case is very similar to that for n=1. To obtain
it one needs only to make some minor modifications on the n=1 algebra.
The R symmetry group for n=2 is $SO(5)$, and so one replaces $T_a$ in the n=1
algebra with $T_{ab}$, ($SO(5)$ generators), a, b run from 1,...,5. 
($\sigma'$)/2 in the n=1 algebra is replaced by $(1/4)(-i/4)
[ \Gamma '_{a}, \Gamma '_{b}]$. R symmetry $SO(5)$ $\Gamma$ matrices are 
primed to prevent confusion with Euclidian $SO(d)$ $\Gamma $ matrices.\foot 
{Part of the factor (1/4) accounts for the over counting in the range of indices
when the generators of $SO(5)$ R symmetry are contracted with the matrix above.
The factor may be checked by noting that the n=2 algebra has 2 n=1 sub-algebras
(because $SO(5)$ contains $SO(4)=SU(2)\times SU(2)$), and then requiring 
that the sub-algebras reduce to the n=1 case derived above.} 

As an example, the last equation in our section on n=1, when adopted to n=2
reads
\eqn\sqsnnncom{ \{ Q'_{i\alpha}, {Q'}_{j}^{\theta \dagger} \} =(\delta_{ij}(i)/2)
[(M_{\mu\nu}
P_{+}\Gamma_{\mu}\Gamma_{\nu})_{\alpha}^{\theta} + 2(P_{+})_{\alpha}^{\theta}D] \  \ +(8/4)(T_{ab}(-i/4)[\Gamma '_a, \Gamma '_b])_i^j (P_{+})_{\alpha}^{\theta}]}
Where it is to be remembered that the $\Gamma '$ matrices have nothing to do with
spacetime $\Gamma$ matrices.

\subsec{ {\bf First Unitarity Restrictions}}

We now derive level one unitarity restrictions on representations of
the superconformal algebra. Some general comments first.
As for conformal algebras, our representation will be irreducible unitary
 modules chosen to be diagonal in the maximal compact subgroup ; 
 Rsym$\times SO(2) \times SO(d)$. States in the module are labeled by 
R symmetry weights, a scaling dimension (Eigenvalue of $D'$) and Lorentz Group
weights. As for the conformal group, the set of states
with a given dimension (eigenvalue under the $SO(2)$) form a finite dimensional,
and in general reducible, representation of $SO(d) \times (R\  \  algera).$ 
Weight vectors in a representation of physical interest always have a lowest 
dimension.
States of this lowest dimension $\e$ host a representation of $SO(d)\times SO(2)\times R \  \ symmetry$ 
that is necessarily irreducible. The lowest (or highest) weights or this finite
dimensional irreducible representation label and completely specify the entire  module. 

Every state in our module may be be obtained by allowing every $Q'$ operator
to act on the lowest weight state at most once, and then permitting 
the conformal group operators  to act on these $Q$s an indefinite 
number of times.
 The superconformal irreducible representation splits into 
a number of different conformal group irreducible representations whose lowest weight states are
all obtained by acting specific combinations of $Q'$s on the Lowest weight state
of the entire module.

We now proceed to derive unitarity constraints on the superconformal modules. We
perform some calculations at the lowest level - in analogy with the calculations 
performed for the conformal group.

\cl{ {\bf d=3}}

Let our states of lowest dimension be denoted by
$|\{t\}m\rangle$ where $\{t\}$ labels $SO(n)$ (R sym ) weights, and $m$ is the 
half integer that labels weights in 
the Lorentz $SO(3)$.
Unitarity demands the positivity of the matrix
\eqn\nmatt{ A_{\nu\{t\}m'j,\mu\{s\}mi} = <\{t\}m'|S'_{j\nu}Q'_{i\mu}|\{s\}m>}
Using \qsncom and $D' = -i\epsilon_0$ , this implies that all negative
eigenvalues of the matrix $B$ have modulus $\leq \e$, where  
\eqn\agna{  B_{\nu\{t\}m'j,\mu\{s\}mi} = (i\delta_{ij}/2)(M_{ab})_{m'm}\Gamma_{a}\Gamma_{b}C)_{\mu\nu} - (i)\delta_{\mu\nu}(I_{ij})_{\{s\}\{t\}} }
Happily $B$ is the sum of two matrices, one of which has only Lorentz indices
and the other of which has only R sym indices, so we may separately 
diagonalize each piece and add the relevant eigenvalues.

The R symmetry piece is diagonalized exactly as we diagonalized $M_{ij}$ when
dealing with the conformal group.
The various eigenvalues of this part of B are\foot{Careful with the order of 
indices - $I$ appears as a transpose} 
$-\half [ c_2(R') +c_2(R) + c_2(vector)]$ 
for various $R'$ occurring in the Clebsch Gordan of decomposition of 
$vector \times R$. 
For the most negative
eigenvalue we pick the $R'$ with the largest Casimir. If R has highest weights ($h_1 ...,h_n$)
then the $R'$ with the largest Casimir has highest wt ($h_1+1, h_2,h_3...,h_n$), 
and so the most negative eigenvalue is simply = $-h_1$ (using \casimir ) .

The Lorentz part is diagonalized similarly. We recognize the matrix expression 
as the matrix element of the operator 2J.S where S is a spin operator, and J 
is an angular momentum operator in the given representation\foot{note M appears as a transpose, and $-M^T$ obeys the same commutation
relations as M}. The lowest eigenvalue of this operator is 0 for $j=0$, or 
$-(j+1)$  otherwise. So we have\foot{These results agree with the 
special cases 
worked out for n=8, $j=0$ by Seiberg, reported in \rseiberg\ , equation 4.1. When
comparing result one must be careful to account for the fact that in \rseiberg\
supercharges were taken to transform in the spinor representation of
$SO(8)$, (possible because of triality)}
\eqn\lt{ \epsilon_0 \geq h_1 \  \ (j=0)}.
\eqn\llt{ \epsilon_0 \geq h_1 + j + 1\  \ (j \neq 0)}
Where $h_1$ is the highest weight of the representation under the R symmetry group
and $j$ the highest weight of the $SO(3)$ Lorentz representation. Note that even 
for $h_1=0$, this is a slightly stronger condition than that obtained from
 conformal invariance alone [\confcondt\ - \confcondttt\ ].

For comparison with the next section it is useful to have a list
of all values of $\e$ (not merely the maximum) for which our $A$ matrix has a 
zero eigen value. These values are 
\eqn\lllt{ \epsilon_0 = j+1-\delta_{0j} + \half(c_2(R') -c_2(R) -(n-1))}
\eqn\llllt{ \epsilon_0 = -j + \half(c_2(R') -c_2(R) -(n-1))}
Where $R'$ runs through all allowed representations. Using appendix 2, we work out all allowed 
values of $\half(c_2(R') -c_2(R) -(n-1))$. They are\foot{ Here, as through
the rest of this paper, slight differences arise between the cases n even and 
n odd, for $d=3$. The cases of interest for physics are n=1,2,4,8.
The case n=1 is easy and has been explicitly worked out in \rheidenreich . Therefore
here, and in the rest of the paper we list formulae only for n even. The 
corresponding formulae for odd n are generally slight modifications. Here, for
instance, n odd admits all the values listed plus $-{n-1 \over 2}$ } 
$h_i + 1-i$ for those 
values of 
i for which $h_{i-1} \neq h_i$, and $-h_i -n +i +1$ for those values of $i$ 
for 
which $h_i \neq 0$, so that if we define
\eqn\cloans{\eqalign{ c^1_{i,n} &= 2-i+j -\delta_{0j} + h_i \cr
 c^2_{i,n} &= 1-i -j + h_i \cr
 c^3_{i,n} &= 2-n+i+j-\delta_{0j} -h_i \cr
 c^4_{i,n} &= 1-n+i -j -h_i \cr}}
we find that our matrix has zero eigenvalues at $\e$ equal to $c^1_{i,n}$ or
$c^2_{i,n}$ for $i$ such that $h_{i-1} \neq h_i$ and $\e$ equal to 
$c^3_{i,n}$ or $c^4_{i,n}$ for i such that $h_{i} \neq 0$.

				\cl{ {\bf d=4}}

Once again we demand the positivity of

\eqn\nmattt{ A_{\nu\{t\}m'n'j,\mu\{s\}mni} = \langle\{t\}m'n'|S'_{j\nu}Q'_{i\mu}|\{s\}mn\rangle}
$\{s\}$ and $\{t\}$ now stand for $U(n)$ weights, $(m, m')$, $(n, n')$ are the 
``$J_z$'' values of our states for the 2$J$s, $J_1$ and $J_2$ that comprise
 the rotation
$SO(4)$. Let $j_1$ be the highest wt of the positive chirality angular momentum $J_1$ and $j_2$ the highest weight of the negative chirality angular momentum $J_2$.
As usual, we process the RHS of \nmattt\ using  \sqfncom . The condition, as for
$d=3$, is of the form $\epsilon_0 \geq$ - (smallest eigenvalue of auxiliary 
matrix). The presence of 2 chiralities introduces a slightly new feature; the 
auxiliary matrix is a direct sum of two matrices, one associated with +ve 
chirality, the second with -ve chirality - we call these two matrices the 
$P_+$ and $P_-$ matrices respectively. Recalling that in the transition 
from $SO(4)$ to $SU(2)\times SU(2)$, $M \cdot M =2( (J_1)^2 + (J_2)^2)$ and being 
careful
about signs, the $P_+$ matrix is (in operator form) $4J_1 \cdot S_1 -2T_{ij}V_{ij} +
 R/2$. Here $V_{ij}$ is the fundamental $SU(n)$ matrix generator in the notation 
of Appendix 1. The $P_-$ matrix is $4J_2 \cdot S_2 -2T_{ij}(V^*)_{ij}+\half R$, where $V^*$
is the $SU(N)$ generator in the antifundamental representation, i.e., 
$[(V^*)_{ij}]_{mn}=-\delta_{in}\delta_{jm}$. 

To compute $2T \cdot V$ we note it is
$(T+V)^2 -T^2 -V^2$, i.e., $C(T\times V) -C(T) -C(V)$. $T\times V$ stands for all 
representations
that can be obtained from $T$ by tensoring with $V$. If $T$ is characterized by
the numbers $\{ r_i\}$ (appendix 1), thinking in terms of a Young Tableaux, 
we see that the only representations that can occur in $T\times V$ are those 
with $r_i$ increased by unity for a single value of $i$ ($=k$ say) and 
 remaining unchanged for all other $i$. The allowed values of $k$ are those for 
which $R_{k-1} \neq 0$ 
(appendix 1). When $T\times V$ couple to give $r'_k=r_k + 1$ the operator 
$(2T \cdot V-\half R)$ 
takes value $(2r_k -{2\Sigma (r_i) \over n} +2 -2k +{R(4-n)\over 2n})$. Computation of $T \cdot V^*$ is 
similar. Coupling with $V^*$ corresponds to adding a box to the dual young 
Tableaux, i.e., removing a box from the Young Tableaux. This corresponds to 
putting $r'_i =r_i-1$ for some $i= k$ st $R_k \neq 0$, and leaving $r_i$ unchanged
for all other values of $i$. When $T\times V^*$ couple to give  $r'_k =r_k-1$ the
 value of ($2T \cdot V^*-R/2$) is
$(-2r_k +{2\Sigma (r_i) \over n} -2n +2k -{R(4-n) \over 2n} )$. Of course $2J \cdot S$ has the usual values
$-(j+1)$ and $j$. Putting this all in we see that our matrix has a zero eigenvalue
at 
\eqn\dloansa{\eqalign{\epsilon_0&=d^1_{nk} = 2j_1 +4 +2r_k -2\Sigma(r_i)/n  -2k +{R(4-n) \over 2n} + (-2 \delta_{j_1 0}) \cr 
\epsilon_0&=d^2_{nk} = -2j_1 +2 +2r_k -{2\Sigma(r_i) \over n}  -2k +{R(4-n) \over 2n} \cr}}
Where $k$ runs over values st $R_{k-1} \neq 0 $ in our representation.
\eqn\dloansb{\eqalign{\epsilon_0&=d^3_{nk} = 2j_2 +2  -2r_k +{2\Sigma(r_i) \over n}  +2k -{R(4-n) \over 2n} -2n 
+ (-2 \delta_{j_2 0}) \cr
\epsilon_0&=d^4_{nk} = 2j_2   -2r_k +{2\Sigma(r_i) \over n}  +2k -{R(4-n) \over 2n} -2n \cr}}
Where $k$ runs over values st $R_k \neq 0$ in our representation.

The unitarity constraint is obtained by requiring that $\epsilon_0 \ge $ the 
the largest of these values. Since $d^1 \ge d^2$, and $d^3 \ge d^4$, we want
the maximum of ($d^1, d^3$). $d^1$ takes its maximum at $k=1$, always an allowed
value. $d^3$ takes its maximum at $k=n$, also always an allowed value. Therefore
our unitarity constraint is 
\eqn\ansfo {\epsilon_0 \geq max(d^1_{n1}, d^3_{nn})}
  
This is consistent with, and almost identical to the conditions derived in 
\rdpresult\ for $d=4$. We will have more to say about this in the next section. 

				\cl{ {\bf d=5}}

The evaluations of level one results in this case is very similar to our 
evaluation in $d=3$, and I will only present answers, obtained using \sqfincom.
 A representation that
transforms under the Lorentz Group $SO(5)$ with highest weights $(h_1, h_2)$
and under the R symmetry group $SO(3)$ with highest weight (`angular momentum
value') k possesses a scaling dimension that obeys the following inequality.

\eqn\ansfo {\epsilon_0 \geq -(c_2(R')-c_2(R)-{5 \over 2}) + 3k}
where $R$ is the $SO(5)$ representation $(h_1, h_2)$, and $R'$ is any representation
that can be obtained from $R$ by tensoring it with the spinor representation. 

The values of scaling dimension at which our $`A'$ matrix develops a zero eigenvalue
are 
\eqn\ansfor{\epsilon_0  =  -(c_2(R')-c_2(R)-{5 \over 2}) + 3k}
\eqn\ansforr{\epsilon_0 =  -(c_2(R')-c_2(R)-{5 \over 2}) - 3k-3
(1 -\delta_{k0}) }
for all $R'$s present in the Clebsch Gordan decomposition of $R$ with the spinor
representation of $SO(5)$. Using appendix 2, 
we write an explicit formula for all possible values of
$(c_2(R')-c_2(R)-{5 \over 2}) $ . 
\eqn\ansforp{c_2(R')-c_2(R)-{5 \over 2} =h_1+h_2}
\eqn\ansforrp{c_2(R')-c_2(R)-{5 \over 2} =h_1-h_2-1}
\eqn\ansforrrp{c_2(R')-c_2(R)-{5 \over 2} =-h_1+h_2-3}
\eqn\ansforrrp{c_2(R')-c_2(R')-{5 \over 2} =-h_1-h_2 -4}
The restrictions put are (derived from the rules in appendix 2)

a) If $h_2 =0$ then the 2nd and 4th of these do not occur.  
b) If $h_2 =h_1=0$ then the 2nd, 3rd and 4th of these do not occur.  
c) If $h_2 =h_1$ then the 3rd of these does not  occur. 
Thus our matrix develops a zero eigenvalue at $\e$ equal to 
\eqn\eloansa{\eqalign{ e^1&=3k-h_1-h_2 \cr
	   e^2&=-3k -3(1-\delta_{k0}) -h_1 -h_2 \cr}}
for all values of ($h_1,h_2$)
\eqn\aloansb{\eqalign{ e^3&=3k-h_1+h_2+1 \cr
	   e^4&=-3k -3(1-\delta_{k0}) -h_1 +h_2+1 \cr}}
for  ($h_1,h_2$) s.t. $h_2 \neq 0$
\eqn\eloansc{\eqalign{ e^5&=3k+h_1-h_2+3 \cr
	   e^6&=-3k -3(1-\delta_{k0}) +h_1 -h_2+3 \cr}}
for ($h_1,h_2$) s.t. $h_1 \neq h_2$
\eqn\eloansd{\eqalign{ e^7&=3k+h_1+h_2+4 \cr
	   e^8&=-3k -3(1-\delta_{k0}) +h_1 +h_2+4 \cr}}
for ($h_1,h_2$) s.t. $h_2 \neq 0$

			\cl{{\bf d=6}}

Once again evaluation is simple, and I only present answers obtained using 
\sqsnncom\ and its partner.
 Our representations 
transform under the Lorentz Group $SO(6)$ with highest weights $(h_1, h_2, h_3)$
and, for n=1,2 under the R symmetry group $SO(3)$, $SO(5)$ with highest weights $k$,
 ($l_1, l_2$).
				
\cl{{\bf n=1}}

The scaling dimension obeys
\eqn\ansfio {\epsilon_0 \geq -(c_2(R')-c_2(R)-{15 \over 4}) + 4k}
where $R$ is the $SO(5)$ representation $\{h_i\}$, and $R'$ is any representation
present in the Clebsch Gordan decomposition of $R$ with the chiral spinor 
representation.

The values of scaling dimension at which the `$A$' matrix develops a zero eigenvalue
are 
\eqn\ansfor{\epsilon_0 =  -(c_2(R')-c_2(R)-{15 \over 4}) + 4k}
\eqn\ansforr{\epsilon_0 =  -(c_2(R')-c_2(R)-{15 \over 4}) - 4k-4 +4\delta_{k0}}
with $R'$ varying over the $SO(6)$ representations spoken of above. We list all
values taken by $-(c_2(R')-c_2(R)-{15 \over 4})$ (derived using appendix 2).
\eqn\bnsfor{-(c_2(R')-c_2(R)-{15 \over 4}) = (-h_1-h_2-h_3)}
\eqn\bnsforr{-(c_2(R')-c_2(R)-{15 \over 4}) =(-h_1+h_2+h_3+2) }
\eqn\bnsforrr{ -(c_2(R')-c_2(R)-{15 \over 4})=(+h_1-h_2+h_3+4) }
\eqn\bnsforrrr{-(c_2(R')-c_2(R)-{15 \over 4})=(+h_1+h_2 -h_3+6) }
The first of these always occurs.
The second occurs if $ h_2-\half \geq |h_3-\half| $.
The third occurs if $h_1 \neq h_2 $.
The fourth occurs if $h_2-\half \geq |h_3+\half|$.

Thus our A matrix develops a zero eigenvalue at $\e$ equal to
\eqn\floansa{\eqalign{ f^{1+}&=4k-h_1-h_2 -h_3\cr
	   f^{1-}&=-4k -4(1-\delta_{k0}) -h_1 -h_2 -h_3 \cr}}
for all values of ($h_1,h_2,h_3$)
\eqn\floansb{\eqalign{ f^{2+}&=4k-h_1+h_2 +h_3+2 \cr
	   f^{2-}&=-4k -4(1-\delta_{k0}) -h_1 +h_2++h_3+2 \cr}}
for  ($h_1,h_2,h_3$) s.t. $|h_3-\half| \leq h_2-\half $
\eqn\floansc{\eqalign{ f^{3+}&=4k+h_1-h_2+h_3 +4 \cr
	   f^{3-}&=-4k -4(1-\delta_{k0}) +h_1 -h_2 +h_3 +4 \cr}}
for ($h_1,h_2,h_3$) s.t. $h_1 - h_2 \geq 1$
\eqn\floansd{\eqalign{ f^{4+}&=4k+h_1+h_2+-h_3+6 \cr
	   f^{4-}&=-4k -4(1-\delta_{k0}) +h_1 +h_2-h_3+6 \cr}}
for ($h_1,h_2, h_3$) s.t. $h_2-\half \geq |h_3+\half| $
 
			\cl{ {\bf n=2}}
The derived bound on scaling dimensions is 
\eqn\ansfit {\epsilon_0 \geq -(c_2(R')-c_2(R)-{15 \over 4}) + 2(l_1+l_2)}
where $R$ is the $SO(5)$ representation $\{h_i\}$, and $R'$ is any 
representation
that can be obtained from $R$ by tensoring it with the chiral spinor representation, the values of the term in the bracket being given by \bnsfor - \bnsforr.

Note that 2($l_1 + l_2) = 4k_1$, where $k_1$ and $k_2$ are the 
$SU(2)\times SU(2)$ highest weights in the $SO(4) = SU(2) \times SU(2)$ subgroup
of $SO(5)$, and so our n=2 results are consistent with the n=1 results.

The values of scaling dimension at which our `$A$' matrix develops a zero eigenvalue
are (subject to restrictions given below)
\eqn\ansfor{\epsilon_0 = g^1_{R'} =  -(c_2(R')-c_2(R)-{15 \over 4}) + 2(l_1+l_2)}
\eqn\ansforr{\epsilon_0 = g^2_{R'} =  -(c_2(R')-c_2(R)-{15 \over 4}) +2(l_1-l_2-1) }
\eqn\ansforrr{\epsilon_0 = g^3_{R'} =  -(c_2(R')-c_2(R)-{15 \over 4}) +2(-l_1+l_2-3) }
\eqn\ansforrrr{\epsilon_0 = g^4 {R'} = -(c_2(R')-c_2(R')-{15 \over 4}) +2(-l_1-l_2 -4) }
The restrictions put are (derived from the rules in appendix 2):
a) If $l_2 =0$ then $g^2_{R'}$, $g^4_{R'}$ no longer yield zero eigenvalue dimensions.
b) If $l_2 =l_1=0$ then $g^2_{R'}, g^3_{R'}, g^4_{R'}$ no longer yield zero eigenvalue dimensions.
c) If $l_2 =l_1$ then $g^3_{R'}$ no longer yields an acceptable zero eigenvalue dimension.

It would be too tedious to explicitly list all 16 possible values of $\e$ 
that result from substituting each of \bnsfor\ - \bnsforrrr\ into each of
\ansfor\ - \ansforrrr\ , but that procedure would generate the list of zero 
eigenvalue scaling dimensions of our algebra. 

\newsec{Lie Super Algebras}

The generators of the Superconformal group, like those of Poincare Supersymmetry,
belong a lie super algebra (LSA), an algebra consisting of two parts, 
the even and the odd such that $even \times even =even, \  \ 
odd \times odd = even, \  \ odd \times even=odd,\  \  even \times odd=odd,$ 
Here $\times$ refers to the commutator / anticommutator between elements 
of the LSA as appropriate. Elements of an LSA are also required to satisfy a 
graded Jacobi identity. 

The theory of Lie super algebras has been studied in detail by a number of 
authors. Simple, finite dimensional, complex Lie Super Algebras have been 
completely classified. They are of two types: Classical Lie Super Algebras, 
and Cartan Lie super algebras. We will describe the 
classification of Classical LSAs in subsection $4.1 .$ In subsection $4.2$ we
will state a result by Kac on the representation of LSAs, and discuss how this
result may be put to use in the task of determining conditions for the unitarity
of representations of superconformal algebras. In subsections $4.4-4.7$ we will
apply the Kac criterion  to the special cases of superconformal 
algebras in d=3,4,5,6 respectively, and complete the task of determining 
unitarity restrictions on the scaling dimensions of representations of 
superconformal algebras.

A good general reference for the theory of LSAs is \rkacreview .

\subsec{{\bf Classical Lie Super Algebras}}

A simple finite dimensional complex LSA G = $ G_0 + G_1 $ is said to be 
classical if the representation of the even part, $G_0$ on the odd part $G_1$,\foot{the representation is formed by taking the commutator of even with odd elements - it is a representation of $G_0$ on $G_1$ because the commutator of an
even and an odd element is always odd}
is completely reducible.\foot{i.e., can be decomposed into the direct sum of blocks 
irreducible representations} It turns out that the even part of a classical 
LSA is always a Lie Algebra.

Classical LSA's may be sub classified as those of type 1, those for 
which the representation of $G_0$ on $G_1$ is reducible, and those of type 2, 
those for which the representation of $G_0$ on $G_1$ is irreducible.

$G_0$ for a type 2 LSA turns out always to be a semisimple Lie Algebra.\foot{A Lie Algebra that has no U(1) factors} Below we list all the 
Classical LSAs of type 2, 
specifying their names, their $G_0$ content, and the representation of 
$G_0$ on $G_1$.\foot{ We use Cartan Notation for Lie Algebras; recall that $A_n$ 
is $SU(n+1)$, $B_n$ is $SO(2n+1)$, $C_n$ is $Sp(n)$, $D_n$ is $SO(2n)$}
\eqn\claslsa{\matrix{
G=G_0+G_1 \qquad & \qquad G_0 \qquad & \qquad G_0\  \  Rep\  \ on\  \ G_1 \cr
& & \cr
B(m,n) \qquad & \qquad B_m+ C_n \qquad & \qquad vector \times vector \cr
D(m,n) \qquad & \qquad D_m+ C_n \qquad & \qquad vector \times vector 
\cr D(2,1,\alpha) \qquad & \qquad A_1+ A_1+A_1 \qquad & \qquad vector \times vector \times vector \cr
F(4) \qquad & \qquad B_3+ A_1 \qquad & \qquad spinor \times vector \cr
G(3) \qquad & \qquad G_2+ A_1 \qquad & \qquad spinor \times vector \cr
Q(n) \qquad & \qquad A_n \qquad & \qquad adjoint\cr &\cr}}
.

Type 1, basic,  Classical LSAs turn out to have $G_0$s that are not in general
semi simple. These algebras 
admit a consistent Z gradation, : i.e., $G_1$ splits into into two pieces
$ G_1 = G_{\bar{1}} + G_{\bar{-1}}$ such that  
$G_{\bar{i}} \times G_{\bar{j}} \subseteq G_{\bar{i}+\bar{j}}$.\foot{ Define $G_{\bar{0}}=G_{0}$. $\times$ as 
usual here denotes the (anti) commutator}.
The representation of $G_0$ on $G_1$ is always reducible for type 1 
algebras; however it turns out that the representations of $G_0$ on 
$G_{\bar{1}}$ and on $G_{\bar{-1}}$ are each irreducible and in fact are 
contragradient\foot{ Two representations are said to be contragradient if
the set of weights in one representation is minus the set of weights in the 
other}. We list all the Basic Classical LSAs of type 1, 
specifying their names, their $G_0$ content, and the representation of 
$G_0$ on $G_{\bar{-1}}$.
\eqn\claslsat{\matrix{
G=G_0+G_1 \qquad & \qquad G_0 \qquad & \qquad G_0\  \ Rep\  \ on\  \ G_{\bar{-1}} \cr
& & \cr
A(m,n) \qquad & \qquad A_m+ A_n+C \qquad & \qquad vector \times vector \times 
 C  \cr
A(m,m) \qquad & \qquad A_m+ A_n \qquad & \qquad vector \times vector \cr
C(n) \qquad & \qquad C_{n-1}+ C \qquad & \qquad vector \times C \cr & \cr}}
C denotes the abelian algebra of complex numbers or its one dimensional 
representation.

There is exactly one LSA of type one that is not basic. For this algebra, the 
representation of $G_0$ on $G_{\bar{-1}}$ and its representation on $G_{\bar{1}}$
 are not contragradient. This algebra is named P(n), its even part is the lie 
algebra $A_n$;  $A_n$ is represented on $G_{\bar{1}}$ as the 
(antisymmetric tensor)*, and on $G_{\bar1}$ as the symmetric tensor.

That completes our listing of the classical LSAs. It will be useful 
for the rest of the paper to have descriptions of some of these algebras in 
terms of supermatrices. We provide this description below. See \rkacreview , 
sec 2.1 for more details.

Consider a graded vector space of even dimension m and odd dimension n. 
Operators on this space may be represented as square (m+n)$\times$(m+n) complex matrices. 
These matrices possess an m $\times $m diagonal block that maps even vectors to
 even vectors,
 and an n $\times$ n diagonal block that maps odd vectors onto odd vectors, as well as 
off diagonal blocks that map odd vectors to even ones and vice versa. 
We endow these matrices with a $Z_2$ grading, defining elements belonging to 
the diagonal blocks to be even elements, and  elements belonging to 
off diagonal block above to be odd elements. We now convert this  
space of matrices into an LSA by defining the LSA product $\times$ 
as the commutator / anticommutator of matrices in the usual fashion. 

Let the space of all $m \times n$ matrices with the $Z_2$ gradation above be 
called  gl(m,n). Endow the base m+n dimensional vector space ( the space on 
which our matrices act) with a bilinear form (`scalar product'), denoted below
by F( , ),
that is symmetric on even elements, antisymmetric on odd elements and ensures
 orthogonality between odd and even elements.\foot{See \rkacreview, Sec 2.1 for 
a matrix realization of this scalar product.}

Define the following sub spaces of gl(m, n).

sl(m, n) = Matrices of zero supertrace belonging to gl(m,n).

osp(m,n) = Matrices A in gl(m,n) s.t. F(Ax,y) = F(x,Ay) for all x,y in the base
 vector space. 

One may identify some of the Classical basic LSAs described behind with these
matrix algebras.\foot{ The case A(m,m) is special below because the 
U(1) factor in the even part of sl(m+1,n+1) becomes the identity
matrix when m=n, and so sl(m+1, m+1) is not a simple LSA. One thus has to mod
it out by identity to obtain the simple LSA A(m,m).} 
$$\eqalign{A(m,n)& = sl(m+1,n+1)\  \ (m\neq n \  \ m,n\geq 0) \cr
	   A(m,m)& = sl(m+1,m+1)/I \  \ (m \ge 0) \cr
	   B(m,n)& = osp(2m+1,2n) \  \ (m\geq0\  \ n > 0) \cr
	   C(n)  & = osp(2,2n)  \  \ (n > 0)    \cr     
           D(m,n)& = ops(2m, 2n) (m\geq 2 ; n \geq 0 ) \cr }$$

Q(n) and P(n) may also be represented as matrices.

Q(n) ($n \geq 2$) is the sub-algebra of A(n,n) which consists of the matrices 
$$\left(\matrix{a&b \cr
			b&a \cr}\right)$$
where a and b are $n\times n$ matrices with tr(b)=0.

P(n) ($n \geq 2$)is the sub-algebra of A(n,n) that consists of matrices
$$\left(\matrix{a&b \cr
			c&-a^T \cr}\right)$$
where a,b,c are $n\times n$ matrices with tr(a)=0 and b=symmetric, 
c=skew symmetric.

This completes our summary of the classification of simple classical LSAs. 
Note that the even part of a Cartan LSA cannot be a lie algebra,
because all representations of a Lie algebra are reductive. 

A couple of definitions we will use ahead. The even roots of 
a classical LSA are the roots of its even component, $G_0$. The odd roots 
of a classical LSA are the weights of the representation of 
$G_0 \  \ on \  \ G_1$. 

The algebras that we will identify as superconformal algebras each possess a
unique bilinear form, i.e., a map ( , )\foot{(a,b) will sometimes be written
as a.b in the next section} : $\CA \times \CA \r C$, which enjoys the following properties. 1) (a,b)=0 if a is even and b is odd or vice versa. 2) 
(a,b)= $(-)^{n}$(b,a), where n=0 if a,b are even, and n=1 if a,b are odd. 3)
(a$\times $b,c) = (a, b$\times$c). For those superalgebras that have a simple
super-matrix realization, a realization of (a,b) is $str(ab)$.\foot{$str$ 
denotes the supertrace, i.e. the trace of the upper diagonal block minus the
trace of the lower diagonal block of a super-matrix}  

Finally a comment on positivity. As in Lie algebra theory it is convenient to 
impose a notion of positivity on the roots of our LSA. Defining positivity
is equivalent to specifying a maximal solvable sub-algebra -the sub-algebra of 
raising operators with positive roots - in our LSA.\foot{a sub-algebra
is said to be solvable if multiplying the sub-algebra with itself a finite 
number of times 
gives zero ; a maximal solvable sub-algebra is called a Borel sub-algebra}. 
All Maximal solvable 
sub-algebras in a $Lie$ algebra are equivalent under similarity transformations,
and so all choices of positivity on a Lie algebra are equivalent. The 
corresponding result is not true for Classical LSAs. The set of inequivalent
conventions of positivity (inequivalent Borel sub-algebras) have been classified
and listed in proposition 1.2 in \rkacmain . It will only be be important for
us to note that not all conventions of positivity give the same results - we 
will have to adopt one on physical grounds ahead.

\subsec{{\bf How Superconformal algebras fit into the classification }}

It is now easy to see why no superconformal algebras exist in $d>6$.
By a superconformal algebra we mean an algebra whose 
even part $G_0$ possesses as a sub-algebra the conformal algebra $SO(d,2)$, 
represented spinorially on the odd part of the algebra, $G_1$.  
Shnider (\rshnider , lemma 1) has shown that if there exists a complex 
superconformal algebra, then there exists a simple one. But we have a list of 
all simple classical  algebras.\foot{ We deal with complex algebras at the moment, so do not distinguish
between $SO(d,2)$ and $SO(d+2)$.} Which classical LSAs have $SO(d)$ sub-algebras? 
The 
LSAs B(m,n) and D(m,n) do, but these are always represented vectorially on 
the odd part of the algebra according to \claslsa\ , \claslsat\ and so are not
superconformal algebras. Other than these,
the only $SO(d+2)$ factor explicit in \claslsa\ , \claslsat\  is in F(4), 
which has a $B_3=SO(7)$ factor. This is indeed represented spinorially on the odd part. Thus F(4) is a superconformal algebra. From \claslsa\ we identify the R symmetry
of this superconformal algebra as $A_1=SU(2)=SO(3)$. The spinors transform in
the vector of $SU(2)$, i.e., the spinor of $SO(3)$. Thus F(4) is the $d=5$, $n=1$ 
superconformal
algebra we have constructed in the previous section. We now have a `reason' for
why the only superconformal algebra in $d=5$ is $n=1$.

	This analysis seems to indicate that there are no superconformal 
algebras except $n=1$ in $d=5$, but that is not quite right, as we have 
omitted to account for isomorphisms between low dimensional lie algebras. 
Recalling
that $SO(5)=Sp(2)=C_2$, and that the spinor representation of the first is the 
vector representation of the second, we immediately deduce that B(m,2) and 
D(m,2) are superconformal algebras in d=3, with R symmetry algebras $B_m=
SO(2m+1)$ and $D_m=SO(2m)$ respectively - these are the $d=3$, n=2m+1 and n=2m 
superconformal algebras we have constructed in the previous section.
Since $SO(6)=SU(4)$ and that the spinor representation of the first
is the vector representation of the second, we deduce that A(3,m) is a 
superconformal algebra in d=4, with R symmetry $A_m + C$, i.e., $SU(m+1) \times
U(1)=U(m+1)$.\foot{Except when m=3, in which case the R symmetry may be either $SU(4)$ or $U(4)$,
depending on whether the algebra is simple or not}These are the $d=4$, n=m+1 
superconformal algebras we have constructed in the previous section.
The case $d=5$ we have dealt with - $SO(7)$ has no isomorphisms, and so there 
exists only the n=1 algebra F(4). $SO(8)$ has no isomorphisms 
either and so it seems that no $d=6$ superconformal algebra exists. 
However $SO(8)$ possesses the magical property of triality - its vector and 
spinor representations are permuted under automorphisms, and so B(4,n)
supplemented by a triality transformation, is the superconformal algebra in
$d=6$, with R symmetry $C_m=Sp(m)$ corresponding to the n=m algebras constructed
in the previous section. Note that under a particular triality automorphism,
all vectors are mapped to spinors of
the a particular chirality. Therefore all $SO(8)$ spinors V in the $d=6$ 
superconformal algebra have the same chirality, as in our construction in the 
previous section.\foot{An $SO(8)$ spinor of positive chirality is composed of
2 $SO(6)$ spinors, one of positive chirality - $Q$ - and one of negative chirality
- $S$ }
 $SO(m)$ for $m > 8$ is not
isomorphic to any other Lie algebra, nor is its vector representation equivalent
to its spinor representation in any way. Hence we conclude that there exist no
superconformal algebras in $d > 6$.

\subsec{{\bf Kac Theory and Unitarity}}

Recall what we know about representations of superconformal algebras. 
Representations are modules of states, that can be built out of a lowest weight
state by successively acting upon it with $ Q'$ raising operators. 
The representation is completely determined by the lowest weight state, which
in turn is specified by a scaling dimension $\e$, the weights of an $SO(d)$ 
representation, and the weights of the R symmetry representation. Group 
commutation relations completely specify the scalar product between different
states in the module, and thus it is not obvious, and indeed not true, that 
the module will have states with positive norm for all possible lowest weight 
states. 

However states at level m,\foot{states obtained from the 
lowest weight state by the application of m $Q'$ operators} have norms 
 that are polynomials in $\e$ of degree $ \leq m$ with positive 
coefficient for the leading power of $\e$. 
Therefore all states have positive norm at large enough $\e$. 
Consider a sequence of representations with fixed R sym $\times$ $SO(d)$ lowest
weights but varying $\e$. At high enough 
$\e$ all states in the module have positive norm.
 As we
decrease $\e$ towards zero, we eventually hit a value, say $\e =a$, at
which a single state $|\psi \rangle$ and all its descendents attain zero norm.
At this value of $\e$ the module 
becomes reducible.\foot {because the set of states modulo zero norm states provide
a representation for the algebra, - as in a gauge theory, where physical states
represent the Lorentz algebra} As $\e$ is lowered below $a$, states develop 
negative norm and the theory becomes non unitary. 

If $| \psi \rangle$ is a level one state then its norm as a function of $\e$ is  
proportional to $(\e-a)$ and that is the end of the story. 
Unitary representations exist for $\e \geq a$ and do not exist
for $\e < a$. If $|\psi \rangle $ is a level 2 state then its norm is proportional
to $(\e -a)(\e - b)$. As $\e$ is lowered below $b$, the set of states generated
from $|\psi \rangle$ regain positive norm, and so these states provide no obstruction to unitarity for $\e \le b$, though other states might.
Again at exactly $\e =b$
the  representation becomes reducible. If $|\psi \rangle$ is a 
level 3 state then its norm is $K(\e-a)(\e-b)(\e-c)$, and the positivity
of this norm leaves open a `window' [c,b] of values of $\e$ that 
could lead to unitary representations ... etc., etc.,. 

The discussion in the paragraph above is useful for the following reason -
Kac (\rkacmain , sec4, theorem 1, part l ) has derived a remarkably simple
criterion for a module representation of a lie superalgebra to become reducible
(i.e., develop a zero norm state ). His theorem, adopted to the case of a lowest
weight module (he used highest weights) states
 { \it A module with lowest weight $\Lambda$ is reducible if and only if 
$(\Lambda ,\rho) = (\Lambda , \alpha)$ where $\alpha$ is an odd root of the algebra with
 zero norm, ( , ) is the non-degenerate bilinear form on the algebra 
introduced in the previous section, and $\rho$ is the difference between the
half sum of positive even roots and the half sum of positive odd roots of the
algebra.}

In the rest of this paper we apply this criterion to the representations of 
d=3,4,5,6 superalgebras, and determine the values (a, b, c ..) of $\e$ that 
(for fixed 
other weights specifying the lowest weight state) give rise to reducible
representations, i.e., zero norm states. 

From the discussion above we see that requiring 
$\e \geq max (a,b,c...)$ is sufficient to guarantee 
unitarity of representations. If, additionally, the zero norm state that forms
as $\e=$ this maximum is a level one state then as this is also 
a sufficient condition for unitarity. This will turn out to be the case
for large enough
representations of the Lorentz $SO(d)$ labeling our lowest weight states.
For small representations of the Lorentz $SO(d)$ algebra (eg $j=0$ in $d=3$), 
it will turn 
out that the state whose norm goes to zero  at the $\e = max(a,b,c...)$ is not
a level one state. In these cases additional argument and calculations will
be employed to almost completely determine necessary conditions for unitarity.

\subsec{{\bf d = 3 }}

	The $d=3, n=n$ superconformal algebra is identified with $osp(n,4)$, and 
so is composed of matrices of the form
$$\left(\matrix{n\times n& n\times 4 \cr
                             4\times n& 4\times 4\cr}\right)$$
Diagonal blocks contain even elements, the off diagonal blocks the 
odd ones. The even part of our algebra is $SO(n) \times SO(5)$. The conformal 
group lives in the $4\times 4$ and the $S0(n)$ R symmetry group in the 
$n\times n$. In the workout below we will assume that $n$ is even - the case n odd
may be worked out analogously.

Conformal generators are identified within the 
$4\times 4$ block as follows (see \sodm\ - \sodkp\ for notation).   
$$S_{ab}=\half\left(\matrix{0&0. \cr
                             0&-(S^{sp}_{ab})^T\cr}\right)$$
Here the $4 \times 4$ matrices $S^{sp}$ are $SO(3,2)$ spinor matrices created 
from the $\Gamma$ matrices given at the beginning of section 2. 

Define 
$$C'=\left(\matrix{0&C \cr
			C&0 \cr}\right)$$
where C is the $2 \times 2$ charge conjugation matrix for $SO(2,1)$, defined
in the d=3 part of section 3. $C'$ obeys $C'^{-1}\Gamma C'=\Gamma^T$ and 
$C'^T=-C'$.
Define $e^i_\alpha$ as the matrix with one in the $(N+\alpha)^{th}$ row, and 
$i^{th}$ column, and zero everywhere else.  Odd elements in the d=3 
superconformal algebra are identified with matrices as 
\eqn\st{V_{i\alpha}=e^i_{\alpha} +i C_{\alpha\theta}(e^i_{\theta})^T}
Where V is defined at the beginning of section 2.

Finally, $S0(n)$ generators are given by standard vector representation 
generators of $SO(n)$ in the n$\times$n block, i.e., $(I_{ij})_{ab}=(-i)
(\delta_{ia}\delta_{jb} -\delta_{ib}\delta_{ja})$ 

One may check that the matrices defined above obey the (anti) commutation 
relations of the $d=3$ superconformal algebra. As usual, we are interested not
in the generators themselves, but in their primed counterparts \newm\ - \newk ,
\qprime , \qprimee . The matrices corresponding to these operators may now 
be constructed from their definitions in the equations just quoted. 

Primed 
operators take on a simpler form in a new basis.\foot{ i.e., on performing a similarity
transformation on our osp(n,4) matrices, $M'=A^{-1}MA$, where A is a matrix
with identity in the $N \times N$ block, zero off diagonal pieces, and matrix
B in the $4 \times 4$ block, with B given by 
$$B={1 \over{\sqrt{2}}}\left(\matrix{1&-i\sigma_0 \cr
			1&i\sigma_0 \cr}\right)$$}

In this basis the two diagonal primed operators are 
(in the $4 \times 4 $ block.)
$$iD'=\half\left(\matrix{I&0 \cr
			0&-I \cr}\right)$$
$$M'_{23}=\left(\matrix{-\sigma_3&0 \cr
			0&\sigma_3\cr}\right)$$
The matrices $I'_{12}, I'_{34} ...$ are made diagonal by the standard $SO(n)$ 
basis transformation, to give $I'_{12}=diag(1,-1,0,0,..)$ etc.. $D', M'_{23}$ 
and these Cartan $Is$ above constitute Cartan sub algebra (CSA) of our algebra.
Linear functionals on this
Cartan sub algebra can always be written as the operation of multiplying the
CSA element with a diagonal matrix (of the CSA form) and taking supertrace
With this understanding we can treat diagonal matrices as linear 
functionals on H, the Cartan sub algebra. Define $e_i$ as the diagonal 
super-matrix with one in the $i^{th}$ $(i \leq n)$ entry, and zero elsewhere. Define
$f_i$ ($i \leq 4$) as the diagonal super-matrix with one in the 
$(n+i)^{th}$ entry, and zero elsewhere. Note $(e_i,e_j)=\delta_{ij}$, $(f_i,f_j)
= -\delta_{ij}$. Define $\alpha_i = (e_{2i-1}-e_{2i})/2$ ;
$\beta_D=-(f_1+f_2-f_3-f_4)/2$ and $\beta_J=-(-f_1+f_2+f_3-f_4)/2$. Note that
$(\alpha_i ,\alpha_j)=\half \delta_{ij}$ and $(\beta_i,\beta_j)=-\delta_{ij}.$ 
\foot{We have chosen the definitions of $\alpha_i$ etc., such that,
$(\alpha_i, H_j)=\delta_{ij}$  etc., (the right hand side is always one or zero).
Therefore $\beta_D$ is a weight vector with positive unit $iD'$ weight, 
and no weight under any other Cartan Generator, etc..}

The even roots of our algebra (the roots of the even part $G_0$) are those
of the conformal algebra
$\pm \beta_D \pm \beta_J$, $\pm \beta_D$, $\pm \beta_J$ (eight in all), and
those connected with the R symmetry $(\pm \alpha_i \pm \alpha_j)$ The odd 
roots are $(\pm \half \beta_D \pm \half \beta_J \pm \alpha_i)$  

	Simple positive roots, may be chosen as
\eqn\ra{r_1=\beta_D-\beta_J}
\eqn\rb{r_2=\beta_J}
\eqn\rc{r_3=\beta_D/2 -\beta_J/2 -\alpha_1}
\eqn\ta{t_1=\alpha_1-\alpha_2}
and so on until
\eqn\tb{t_{{n\over 2}-1}= \alpha_{n/2-1}-\alpha_{n/2}}
This choice of positivity is partly dictated by the requirement that all 
positive roots, acting on D, yield a positive number. It corresponds 
intuitively to the ordering $\beta_D \ge \beta_J \ge \alpha_1 ... \ge \alpha_{n/2}$.
The half sum of even positive roots, and odd positive roots 
$\rho_0$, and $\rho_1$ are given  by 
\eqn\rhot{\rho_0={3 \over2}\beta_D +\half \beta_J +\sum_{i=1}^{n/2} \half \alpha_i(n-2i)}
\eqn\rhoot{\rho_1=\half n \beta_D}
\eqn\rhooot{\rho = \rho_0 - \rho_1 = {(3-n) \over 2} \beta_D + \half \beta_J
+\half \sum_{i=1}^{n/2} \alpha_i(n-2i)}  
One may now take the inner product of $\rho$ with odd positive roots. 
Use 
\eqn\oddroo{ (\half \beta_J+\half \beta_D).\rho =-1+{n \over 4}}
\eqn\oddrooo{ (\half \beta_D -\half \beta_J).\rho =-\half+{n \over 4}}
\eqn\oddd{\alpha_i.\rho = {(n-2i) \over 4}}
The odd positive roots of zero norm  are $\half \beta_D \pm \half \beta_J \pm \alpha_i$

If our lowest weight multiplet has dimension $\epsilon_0$ and lowest $SO(n)$ 
wts = $(-h_1, -h_2, -h_3  -h_{n/2})$, and has $SO(3)$ lowest wt = $-j$, then 
the
lowest weight in the module is 
\eqn\lws{ \Lambda =-h_1 \alpha_1 -h_2 \alpha_2 -...-h_{n/2} \alpha_{n/2} +d 
\beta_D -j \beta_j} The 
reducibility conditions are ($\Lambda$, root) = ($\rho$, root). 
These conditions, for the odd positive zero norm roots above are

\eqn\cloansn{\eqalign{ c'^1_{i,n} &= 2-i+j  + h_i \cr
 c'^2_{i,n} &= 1-i -j + h_i \cr
 c'^3_{i,n} &= 2-n+i+j -h_i \cr
 c'^4_{i,n} &= 1-n+i -j -h_i \cr}}

These are precisely the conditions \cloans\ without the restrictions on the 
range of $j$, and without delta function exceptions for the case $j=0$.

The full unitarity restrictions for $d=3$ are now easy to work out. For 
$j\neq 0$
the largest $c'$ is $c'^1_{1,n} = c^1_{1,n} = 1+j+h_1$. Since a 
level one state attains zero norm at this dimension, we conclude that 
$\e \geq 1+j+h_1$ is the necessary and sufficient condition for unitarity of
our module. For $j=0$ a level one state does not attain zero norm at the largest
$c'$, but does at the second largest $c'$ ($c'^2_{1,n}=c^1_{1,n}$). Explicit 
calculation that we do not reproduce here shows that at level 2 there is a state
with norm $(\e - c'^1_{1,n})(\e - c'^1_{2,n})$. Therefore unitarity allows only
$\e \geq 1+h_1 $ or $\e = h_1$ (remember $j=0$). Actually more work is needed 
in principle to show that the isolated value of $\e$ is allowed at all levels, 
but we circumvent that procedure by noting that, independently, for $j=0$ we 
have very
good reason to believe a unitary representation at this value of $\e$ 
does exist.\foot{See, for instance \rseiberg\ , section 4} 

The calculations above have been presented for the case n=even, but the 
generalization to n=odd is is a simple matter. However we do not even need to
perform this generalization; the only odd n of interest to us is n=1 and this
special case is rather simple, and has been explicitly worked out (by hand) in
a nice paper by Heidenreich \rheidenreich - I list his answers  here - unitary
representations are obtained for $\e \geq \half$ if $j=0$, and $\e \geq j+1$ for 
$j > 0$. Inspection shows that this is condition \scalarfin\ for $j=0$ and 
condition \llt\ for $j>0$.    

\subsec{{\bf d = 4}}

This subsection is included in this paper only for completeness, everything
in this subsection may be found in \rdpresult , \rdpdetails , \rdpreview .

 A(3,n-1) consists of matrices of the form
$$\left(\matrix{4\times 4& 4\times n \cr
                             n\times 4& n\times n\cr}\right)$$
Diagonal blocks even elements, the off diagonal blocks are 
odd ones. The conformal group basically
lives in the $4\times 4$ and the $SU(n)$ R symmetry group in the $n\times n$. 
The U(1) lives in both diagonal blocks. Specifically, conformal generators
 are identified within the $4\times 4$ block as follows.

$$J'^2_i=\half\left(\matrix{0&0. \cr
                             0&\sigma_{i}.\cr}\right)$$
$$J'^1_i=\half\left(\matrix{\sigma_{i}&0. \cr
                             0&0\cr}\right)$$
$$iD'=\half\left(\matrix{I&0. \cr
                             0&-I\cr}\right)$$
$$P'_{\mu}=\left(\matrix{0&\sigma_{\mu} \cr
                             0&0\cr}\right)$$
$$K'_{\mu}=\left(\matrix{0&0 \cr
                             \sigma_{\mu}&0\cr}\right)$$
Where we have defined $\sigma_{\mu} =(I,\sigma_i)$ Define also $e^i_\alpha$ 
as the
matrix with 1 in the $\alpha^{th}$ row and $(4+i)^{th}$ column and zeroes everywhere 
else. Further define $f^i_{\alpha}$ as the matrix with 1 in the (2+$\alpha)^{th}$
row and $(4+i)^{th}$ column. We make identification of odd elements as
\eqn\qq{Q'^{\dagger}_{i\alpha}=e^i_{\alpha}}
\eqn\q{Q'_{i\alpha}=(f^i_{\alpha})^T}
\eqn\sss{S'_{i\alpha}=\epsilon_{\alpha\beta}f^i_{\beta}}
\eqn\ss{S'^{\dagger}_{i\alpha}=\epsilon_{\alpha\theta}(e^i_\theta)^T}
Finally, R charge and $SU(n)$ generators are identified as 
$$R={1\over {4 \over n}-1}\left(\matrix{I&0 \cr
   				0&{4I \over n} \cr}\right)$$
$$T_a=\left(\matrix{0&0 \cr
   				0&-T^T_a \cr}\right)$$
(The matrices above are $ (4+n) \times (4+n)$ )

The Cartan Sub Algebra of our matrix algebra consists 
of the set of supertraceless diagonal matrices. Linear functionals on the CSA
can always be written as the operation of multiplying the
CSA element with a diagonal\foot{and supertraceless for uniqueness - this will not
be important for us we will never use uniqueness} matrix and taking supertrace.
Define the matrices $e_i$ $(i=1,...,4)$ and $f_j$ $(j=1,...,n)$ as in the previous 
subsection.
The even roots of our algebra are $(e_i-e_j)$ and $(f_i -f_j)$. 
The odd roots of our algebra are $(e_i + f_i)$ . 

	Choose a positivity convention that ensures that the state of 
lowest weight in our module has also the lowest scaling  
dimension in the module (as we require physically) i.e., a positivity 
convention that ensures that positive roots acting on $D'$, give a 
positive number. We choose as simple positive roots
\eqn\pra{\gamma_1=(e_1-e_2)}
\eqn\prb{\gamma_2=(e_4-e_3)}
\eqn\prc{\gamma_3=(e_2+f_1)}
\eqn\prd{\gamma_4=(-e_4-f_N)}
\eqn\prn{\gamma_{4+k}=(-f_k + f_{k+1})}
This leads to positivity rules for roots with dictionary ordering and 
$e_1 \ge e_2 \ge -e_3 \ge -e_4 \ge -f1 \ge -f_2 \geq ...\geq  -f_N$.

The half sum of even positive roots, $\rho_0$ and the half sum of odd positive
roots $\rho_1$ and their difference $\rho = \rho_0 - \rho_1$ are.
\eqn\rhef{\rho_0=\half [\sum_{i=1}^n (n+1-2i)f_i] +\half(3e_1 + e_2 -3e_3 -e_4
)}
\eqn\rhof{\rho_1={n \over 2}[e_1 +e_2 -e_3 -e_4]}
\eqn\rhf{\rho = {3-n \over 2}e_1 +{1-n \over 2}e_2 +{n-3 \over 2}e_3 +{n-1 \over 2}e_4 -\half \sum_{i=1}^n (n+1-2i)f_i }

One may now compute the dot product of the odd positive roots with $\rho$. 
\eqn\raf{(e_1+f_i).\rho = 2-i}
\eqn\rbf{(e_2+f_i).\rho = 1-i} 
\eqn\rcf{(-e_3-f_i).\rho = -(n-1-i)}
\eqn\rdf{(-e_2-f_i).\rho = -(n-2i)} 

The lowest weight state in our module may be written as
\eqn\lwf{\eqalign{\Lambda &= {\e \over 2}(e_1+e_2-e_3-e_4) -j_1(e_1-e_2) 
+j_2(e_3-e_4) \cr +
& {(4-n) \over n}{R \over 8}(e_1+e_2+e_3+e_4-f_1-f_2-f_3-...-f_N) +(-\sum_{i=1}^kf_i )r_k \cr}}
In the equation above, $j_1$ and $j_2$ are the (half) integers that label the 
2 $SU(2)$s in the Lorentz group.\foot{The signs in our parameterization above are 
slightly subtle. For inst, $J_2^3$, the z component of $J_2$ is a negative 
member of the Cartan Sub Algebra (because $e_3-e_4$ is negative), and so the
lowest weight state in a $J_2$ multiplet is the state with $m=+j_2$. Therefore
we have chosen our parameterization above such that $\Lambda(J_2^3)=j_2$. On 
the other hand $J_1^3$ is a positive member of the Cartan sub-algebra, and 
so the state with the lowest weight in a $J_1^3$ multiplet is one with $m=-j_1$
and so our parameterization has $\Lambda(J_1^3)=-j_1$. The R charge and $iD'$ 
are 
what they are (have a single given value in the lowest weight sector), 
and so $\Lambda(D)=\e$, $\Lambda(R)=R$ with no funny minus signs. The 
sign behind $r_k$ is similarly deduced.}

The reducibility condition is
$(\Lambda -\rho)\cdot \alpha$ =0 for some odd
root $\alpha$ which has zero norm (i.e., $\alpha \cdot\alpha=0$). All odd
roots of this algebra have zero norm. Using \raf\ -\rdf\ and eqn \lwf, 
 a representation is
reducible iff 
\eqn\done{\epsilon_0=d'^1_{nj} = 2j_1 +4 +2r_j -{2\Sigma(r_i) \over n}  -2j +{R(4-n) \over 2n} }
\eqn\dtwo{\epsilon_0=d'^2_{nj} = -2j_1 +2 +2r_j -{2\Sigma(r_i) \over n}  -2j +{R(4-n) \over 2n} }
\eqn\dthree{\epsilon_0=d'^3_{nj} = 2j_2 +2  -2r_j +{2\Sigma(r_i) \over n}  +2j -{R(4-n) \over 2n} -2n }
\eqn\dfour{\epsilon_0=d'^4_{nj} = 2j_2   -2r_j +{2\Sigma(r_i) \over n}  +2j -{R(4-n) \over 2n} -2n }
Where $j$ runs from 1 to n.

These are exactly eqns \dloansa\ - \dloansb\ without the restrictions on the 
range of j and without the special provisos for the cases $j_1=0$
or $j_2=0$. 

Unitarity conditions for $d=4$ follow. A sufficient condition for unitarity 
is $\epsilon_0 \geq 
max(d'^1_{nj},d'^2_{nj}, d'^3_{nj}, d'^4_{nj})$. Unless either 
$j_1$=0 or $j_2$ =0, the objects $d'$ in \done - \dfour are identical to 
the quantities $d$ in \dloansa\ and \dloansb, and so   
(by now familiar logic) this condition is both
necessary and sufficient condition for positivity of norm in our module.

If, on the other hand, either $j_1=0$ or $j_2=0$ then $d \leq d'$ and the 
necessary condition for norm positivity, \ansfo, is potentially weaker than 
the 
sufficient condition above (obtained, notice,  by replacing $d$ with $d'$ in 
\ansfo ). Actually, if 
$j_1= 0;j_2\neq 0$ and $d'^3_{nn} \ge d'^1_{n1}$, or if $j_1\neq 0;j_2= 0$ and 
$d'^3_{nn} \le d'^1_{n1}$ then once again necessary and sufficient conditions 
coincide, both being given by \ansfo. 

However, if $j_1= 0;j_2\neq 0$ and
 $d'^3_{nn} \le d'^1_{n1}$, then the necessary condition for positivity of 
norm is $\epsilon_0 \geq max (d^1_{n1}=d'^2_{n1} , d'^3_{nn})$ and the sufficient condition 
$\epsilon_0 \geq d'^1_{n1}$ - these do not coincide. So there is a 
window of values of $\epsilon_0$, i.e., the interval $[max(d'^2_{n1},d'^3_{nn}),d'^1_{n1})]$, 
within which
the representation may or may not be unitary, consistent with our derivations
so far. Explicit calculations \rdpreview , show 
that a state with norm proportional to 
$(\epsilon_0-d'^1_{n1})(\epsilon_0-d'^2_{n2})$ occurs at level 2. This says
that if $d'^3_{nn} > d'^2_{n1}$, then no value in the window above is permitted
by unitarity. However if $d'^3_{nn} \leq d'^2_{n1}$, then the isolated
value $\e = d'^2_{n1}$ is not ruled out by any calculation yet reported.
Further calculations reported in the reference just 
cited confirm that in this case
$\e=d'^2_{n1}$ is indeed an allowed value.
Thus in this case the unitarity restriction is then that $\e$ obey $\e \geq d^1_{n1}$ or that
$\e = d'^2_{n1}$. Unitarity constraints for the case $j_2 =0$ are derived 
analogously. All final results are listed in section 5.      

\subsec{{\bf d=5}}
Working out the reducibility conditions for the superconformal algebra in $d=5$,
though simple in principle, is slightly complicated by the fact that F(4) has
no simple matrix representation, and so ( , ) cannot simply be implemented as
a supertrace. Actually an explicit form for ( , ) is known - it is the 
killing form of the algebra ( see \rnahm ) but it seems a bit pointless to 
go through the bother of performing this evaluation, as experience with the 
easier 
cases - $d=3,4,6$ - shows how the reducibility criteria are related to the
dimensions at which level one states attain zero norm. I therefore simply 
conjecture that representations of the d=5 superconformal algebra are 
reducible iff $\e$ is one of 
\eqn\eloansn{\eqalign{ e'^1&=3k-h_1-h_2 \cr
	   e'^2&=-3k -3 -h_1 -h_2 \cr
	  e'^3&=3k-h_1+h_2+1 \cr
	   e'^4&=-3k -3 -h_1 +h_2+1 \cr
	   e'^5&=3k+h_1-h_2+3 \cr
	   e'^6&=-3k -3 +h_1 -h_2+3 \cr
	   e'^7&=3k+h_1+h_2+4 \cr
	   e'^8&=-3k -3 +h_1 +h_2+4 \cr}}
With this assumption we now go on to deduce unitarity bounds for d=5.
Consider Lorentz representations such that 

a) {\bf $h_2 > zero$} . Examples of such representations are the spinor, the 
antisymmetric tensor, and the Rarita Schwinger representation.
In this case the largest $e'$ ; $e'^7$ is = $e^7$, i.e., a dimension at which
a level one state attains zero norm. Therefore representations are unitary
if and only if $\e \geq 3k +h_1 +h_2 +4$. The bounds are $3k+5$, $3k+6$ and 
$3k+6$
for the three examples above.

b) {\bf $h_2=0, \  \ h_1 \neq 0$}. Examples of such representations are the vector
and the symmetric traceless tensor. The two largest $e'$ values are
$e'^7$ and $e'^5$ in that order. No state attains zero norm at $\e=e'^7$ 
at level one (see the condition below \eloansd), but that does happen at
$\e=e'^5 =e^5$ (see the condition below \eloansc). Thus this case is similar 
to the $j=0$ case for $d=3$, that is it is sufficient that 
$\e \geq 3k +h_1 + 4$ for unitarity, but the isolated value $\e = 3k+h_1+3$
may also be permitted by unitarity. No other value of $\e$ could possibly be 
allowed. We have performed some calculations at levels higher than one and 
have not managed to rule out this possibility and so conjecture that this 
isolated value of $\e$ indeed does permit a unitary representation. 
The isolated
value occurs at dimension $(3k+4+h_1)$ and $(3k+5+h_1)$ for the two examples cited above.
The `continuum' of allowed values begins at $(3k+5+h_1)$ and $(3k+6+h_1)$ for these two
examples.

c) {\bf $h_1=0, \  \ h_2=0$} . This is the (Lorentz) scalar representation. The four 
largest $e'$ values are $(3k+4)$, $(3k+3)$, $(3k+1)$, $3k$. Only the last
of these is a dimension at which a state attains zero norm at level 1. 
On general grounds
the first window between $(3k+3)$ and $(3k+4)$ does not permit unitary 
representations except possibly at the lower bound. I have diagonalized the 
equivalent of matrix A in \nmatt\ at level 2 (but for the case d=5, and 
with 2 Qs and 2 $Q^{\dagger}s$ sandwiched between states). At this level 
there appears a state with norm $=(\e-(3k+3))(\e-3k)$. This rules out all
values of $\e < 3k+3 $ other than possibly $\e = 3k$. Therefore the only 
possible values scaling dimension permitted by unitarity are $\e \ge 3k+4$
and possibly $\e = 3k$, $\e = 3k+3$. No norm calculation I have performed has
produced a result that forbids these isolated values : I conjecture that 
unitary representations exist at these two isolated values of $\e$.
Proving this conjecture would require one to check the norms of states at all
levels (upto 8) (a task I have not the energy for), or the use of some clever
trick.

\subsec{{\bf d=6}}

	The $d=6$, n=1;2 superconformal algebra is identified with $osp(8,2;4)$, and 
so is composed of matrices of the form
$$\left(\matrix{8\times 8& 8\times x \cr
                             x\times 8& x\times x\cr}\right)$$
Where x=2,4 for n=1,2.

Diagonal blocks contain the even elements, the off diagonal blocks the 
odd ones. The even part of our algebra is $SO(8) \times SO(2)$ ; $SO(5)$.
The conformal group lives in the $8\times 8$ and the $SO(3);SO(5)$ R symmetry 
group in the $x\times x$. 

Conformal generators are identified within the 
$8\times 8$ block as follows (see \sodm\ - \sodkp\ for notation).   
$$S_{ab}=\half\left(\matrix{-(S^{sp}_{ab})^T&0 \cr
                             0&0 \cr}\right)$$
Here the $8 \times 8$ matrices $S^{sp}$ are the positive chirality projections
of $SO(3,2)$ spinor matrices created from the $\Gamma$ matrices given at the 
beginning of section 2. 

 $S0(3);SO(5)$ generators are given by the negative transpose of their
spinor representations in the $x \times x$ block.

The odd elements of the algebra may be identified with off diagonal matrices,
as in the case of d=3. 
Define $e^i_\alpha$ as the matrix with one in the $(\alpha)^{th}$ row, and 
$(8+i)^{th}$ column, and zero everywhere else.  Odd elements in the d=6 
superconformal algebra are identified with matrices as 
\eqn\st{V_{i\alpha}=A C''_{ij}e^j_{\alpha} +B C'_{\alpha\theta}(e^i_{\theta})^T}
Where V is defined at the beginning of section 2, $C'$ and $C''$ are the charge
conjugation matrices appropriate to $SO(6,2)$ and $SO(3);SO(5)$ spinor 
representations 
respectively and A and B are constants whose value does not concern us.

One now determines primed operators using their definitions. On making the 
the basis change that block diagonalizes $iD'$ and $M'_{ab}$ 
we obtain that $SO(8)$ Cartan algebra elements are represented within the 
$8 \times 8$ block by the matrices
$$ M'_{12} = {1 \over 2} diag(-1,-1,+1,+1,-1,-1,+1,+1)$$
$$ M'_{34} = {1 \over 2} diag(-1,+1,-1,+1,-1,+1,-1,+1)$$ 
$$ M'_{56} = {1 \over 2} diag(-1,+1,+1,-1,+1,-1,-1,+1)$$
$$ iD' = {1 \over 2} diag(+1,+1,+1,+1,-1,_1,-1,-1)$$ 
In the case n=1, the CSA of the superalgebra is spanned by the 4 elements above
and 
$$T_3=-\sigma_3 /2$$ within the $2 \times 2$.
In the case n=2, the CSA of the superalgebra is spanned by the 4 elements above and the following matrices within the $4 \times 4$ 
$$T_{12} = {1 \over 2}diag(-1,-1,+1,+1)$$
$$T_{34} = {1 \over 2}diag(-1,+1,-1,+1)$$

Linear functionals on the CSA can be written as diagonal matrices as behind; 
define $e_i$ (i=1..8) and $f_j$ (j=1..x) as in the $d=3$ workout.
Define
$$ \alpha_4 = {1 \over 4}(-e_1 -e_2 +e_3 +e_4 - e_5 -e_6 +e_7 +e_8)$$
$$ \alpha_3 = {1 \over 4}(-e_1 +e_2 -e_3 +e_4 - e_5 +e_6 -e_7 +e_8)$$
$$ \alpha_2 = {1 \over 4}(-e_1 +e_2 +e_3 -e_4 + e_5 -e_6 -e_7 +e_8)$$
$$ \alpha_D = {1 \over 4}(+e_1 +e_2 +e_3 +e_4 - e_5 -e_6 -e_7 -e_8)$$
In the n=1 case we also define
$$ \beta = f_1-f_2 $$
In the n=2 case we define
$$ \beta_1 = {1 \over 2}( f_1 + f_2 -f_3 -f_4)$$
$$ \beta_2 = {1 \over 2}( f_1 - f_2 +f_3 -f_4)$$

The definitions above have been chosen such that, for instance, 
$\alpha_4(M'_{12}) =1, \alpha_4$(any other element of CSA) =0, etc..
Note$ (\alpha_i . \alpha_j) = {1 \over 2}\delta_{ij}$ and that 
$\alpha s $ are orthogonal to $\beta  s$. Note also  
$(\beta , \beta)=-2$ (relevant to n=1) and $(\beta_i , \beta_j) =-\delta_{ij}$
(relevant to n=2).

The even roots of our algebra (the roots of the even part $G_0$) are those
of the conformal algebra
$\pm \alpha_i \pm \alpha_j$, (24 in all), and
those connected with the R symmetry  $\pm \beta $ for n=1, and 
$\pm \beta_1 \pm \beta_2; \pm \beta_1 ; \pm \beta_2 $ for n=2. 
The odd roots for n=1 are
$\pm \half \alpha_D  \pm \half \alpha_1 \pm \half \alpha_3 \pm \half 
\alpha_4 \pm \half \beta $, (16 in all; the number of plus signs behind 
$\alpha s$ is restricted to be even) 
and for n=2
$(\pm \half \alpha_D \pm \half \alpha_1  \pm \half \alpha_3 \pm
\half \alpha_4 \pm \half \beta_1 \pm \half \beta_2 $, 
(32 in all; the number of 
plus signs behind $\alpha s$ is restricted to be even)

We choose a set of simple positive roots that 
enforces positivity rules that correspond to dictionary ordering with
$\a_D > \a_1 > \a_2 > \a_3 > \b $ for n=1 and 
$\a_D > \a_1 > \a_2 > \a_3 > \b_1 > \b_2 $ for n=2.
The half sum of even positive roots, and odd positive roots 
$\rho_0$, and $\rho_1$ are given  by (n=1)
\eqn\rhoso{\rho_0 = 3\alpha_D +2\alpha_1+\alpha_2+\half \beta }
\eqn\rhooso{\rho_1=2\a_D}
\eqn\rhoooso{\rho = \rho_0 - \rho_1 = \a_D +2\a_1 +\a_2 +\half \b} 
and, for n=2
\eqn\rhost{\rho_0 = 3\alpha_D +2\alpha_1+\alpha_2+{3\over 2}\beta_1 +{1\over 2}
\b_2}
\eqn\rhoost{\rho_1=4\a_D}
\eqn\rhooost{\rho = \rho_0 - \rho_1 =- \a_D +2\a_1 +\a_2 +{3\over 2}\b_1 
+{1\over 2}\b_2 }  

Let the lowest weight multiplet have dimension $\epsilon_0$, lowest $SO(8)$ 
wts  $(-h_1, -h_2, +h_3 )$\foot{Note that the lowest weights in our module have $h_3$ appearing with a positive 
sign. We have chosen this strange looking convention to ensure that the 
representation has highest weights $(h_1,h_2,h_3)$ as was the case in our 
work out in section 3. For instance, a chiral spinor that has highest weights
$(\half ,\half ,\half )$, but lowest weights $(-\half ,-\half ,\half )$.}  and $SO(3);SO(5)$ lowest wts = -j, ; 
$(-l_1 , -l_2)$ . The
lowest weight in the module is then (n=1)
\eqn\lwssa{ \Lambda =  \e \a_D  -h_1 \alpha_1 -h_2 \alpha_2 +h_3\alpha_3
-k\beta } 
and (n=2)
\eqn\lwssb{ \Lambda =  \e \a_D  -h_1 \alpha_1 -h_2 \alpha_2 +h_3\alpha_3
-l_1\beta_1 -l_2 \beta_2}
The 
reducibility conditions are ($\Lambda$, root) = ($\rho$, root). 
These conditions, for the odd positive zero norm roots are (n=1) $\e$ is one of
\eqn\cloansnsa{\eqalign{ f'^{1\pm} &= -h_1-h_2-h_3-2 + \pm(-4k-2) \cr
 f'^{2\pm} &= -h_1+h_2+h_3 + \pm(-4k-2) \cr
 f'^{3\pm} &= +h_1-h_2+h_3 +2 + \pm(-4k-2) \cr
 f'^{4\pm} &= +h_1+h_2-h_3 +4 + \pm(-4k-2) \cr}}
These are exactly \floansa\ - \floansd\ without the delta function provisos and 
the restrictions in those equations.

The reducibility conditions for n=2 are $\e$ is one of
\eqn\cloansnsb{\eqalign{ f'^{1\pm\pm} &= -h_1-h_2-h_3-4+ \pm(-2l_1-3)+
\pm(-2l_2 -1) \cr
 f'^{2\pm\pm} &= -h_1+h_2+h_3 -2 + \pm(-2l_1-3)+\pm(-2l_2 -1)  \cr
 f'^{3\pm\pm} &= +h_1-h_2+h_3  +\pm(-2l_1-3) \pm(-2l_2 -1) \cr
 f'^{4\pm\pm} &= +h_1+h_2-h_3 +2 +\pm(-2l_1-3)+\pm(-2l_2 -1) \cr }}
These are precisely the conditions that follow from \ansfor\ - \ansforrrr\
and \bnsfor\ - \bnsforrrr\ without the restrictions on applicability present in 
those equations.

We now analyze the conditions under which unitary representations of these 
algebras exist. We perform the analysis for the case n=1; the n=2 case is 
extremely similar. All results are listed explicitly in the section 6.

a) {\bf $h_2 -{1\over 2} \geq |h_3+{1\over 2}|$} . Examples of such 
representations 
are the antichiral spinor, the antisymmetric tensor, the antichiral Rarita
Schwinger representation and the antiselfdual antisymmetric 3 tensor. 
In this case the largest $f'$ ; $f'^{4--}$ is a dimension at which
a level one state attains zero norm. Therefore representations are unitary
if and only if $\e \geq 4k +h_1 +h_2 -h_3 +6$. The bounds are 
$4k$ + y, where y =7.5 , 8 , 8.5 and 9 for the four examples above.

b) $h_1-h_2 \geq 1$ $and$ the condition of part a) is not met.
Examples of such representations are the vector,
the symmetric traceless tensor and the chiral Rarita Schwinger representation.
 The two largest $f'$ values are
$f'^{4--}$ and $f'^{3--}$ in that order. No state attains zero norm at 
$\e=f'^{4--}$ 
at level one (see below \bnsforrrr\ ), but that does happen at
$\e=e'^{3--} =e^{3--}$ (see below \bnsforrr\ ). Thus this case is similar 
to the $j=0$ case for $d=3$. It is sufficient that 
$\e \geq 4k +h_1 +h_2 -h_3 +6 $ for unitarity, but the isolated value
 $\e = 4k +h_1 -h_2 +h_3 +4$
may also be permitted by unitarity. No other value of $\e$ could possibly be 
allowed. We have performed some calculations at levels higher than one and 
have not managed to rule out this possibility and so conjecture that this 
isolated value of $\e$ indeed does permit a unitary representation. 
Isolated values occur at $\e=4k+y$ where y=5, 6, 5.5 for the 3 examples 
above.
The continuum of allowed values begins at y=7, 8, 7.5 in these examples.

c){\bf $h_1=h_2=h_3=h \neq 0$ }. Examples are the chiral spinor, and the selfdual
antisymmetric 3 tensor. In this case a level one state attains zero norm only
at the 3rd largest of the $f' s$, i.e., at $f'^{2--}=2 +h +4k =r (defn)$. 
The largest $f'$ is $f'^{4--}=6+h+4k$, a value we define as $t$, and the
second largest is $f'^{3--}=4+h+4k=s (defn)$. On general grounds, unitarity 
permits representations with $\e \geq t$. Unitarity forbids representations
with $\e < r$, and with $\e$ in the range $(s,t)$. The calculations I have
performed at levels higher than one have not managed to rule out the window
$[r,s]$ . However, in the next section I will present an explicit construction
of unitary representations with $\e=r$ and $\e=s$.
So in summary, short unitary representations exist for $\e = t,r,s$. Unitary
representations also exist for $\e>t$. Unitary representations with 
scaling dimensions in the range $(r,s)$  may or may not exist\foot{It may be a relevant 
observation that at $k=0$, $2+h=s$ is 
precisely the scaling dimension of the free conformal scalar field 
with $h_1=h_2=h_3=h$ as derived in section 2. This
fact perhaps hints that unitary representations exist in the full interval
[4k+2,4k+4]}. 

d) {\bf $h_1= h_2=h_3=0$ }  . This is the Lorentz scalar representation. 
The four 
largest $f'$ values are $(4k+6)$, $(4k+4)$, $(4k+2)$, $4k$. 
Only the last
of these is a dimension at which a state attains zero norm at level one.
As usual
the first window $(4k+4, 4k+6)$ does not permit unitary representations. Explicit 
calculations at level 2 show the existence of a state with norm proportional 
to $(\e - (4k+2))(\e -4k)$. Therefore the only allowed values of scaling 
dimension are $\e \geq 4k+6$ and possibly $\e$ in the interval 
$[4k+2 , 4k+4]$ and the isolated value $\e=4k$.In the next section I present 
a construction of short representations with scaling dimension 
$\e=4k, 4k+2, 4k+4$. The calculations I have
performed at levels higher than one have not managed to rule out the window
$(4k+2,4k+4)$ . So in summary, short unitary representations exist for 
$\e=4k, 4k+2, 4k+4, 4k+6$. Unitary representations do exist for $\e >4k+6$ and
may or may not exist for $\e$ in the interval $(4k+2,4k+4)$\foot
{It may be a relevant observation that at k=0, 2 is precisely
the scaling dimension of the free conformal scalar field as derived in section 2. This
fact perhaps hints that unitary representations exist in the full interval
[4k+2,4k+4]}. 
 
\newsec{Examples and Applications}

We are interested primarily in the unitary representations of superconformal
algebras on the space of local operators in a superconformally invariant 
quantum field theory. Three such theories are the (intrinsic) theories on the 
world volumes of the 
$M_2,M_5$ and $D_3$ brane. In the next three subsections we
review what is known about some of the primary operators of these theories and
see how this fits with the results obtained in previous sections. In 
subsections 5.4 and 5.5 I proceed to explicitly 
construct all the 
short representations of the $d=6$ algebras whose existence was left open in 
the previous section. The construction uses the oscillator method.

\subsec{{\bf The World Volume Theory of the $M_2$ brane}}

Consider the $d=3$ $N=8$ superconformal field theory on the world volume of 
a single $M_2$ brane. The only matter 
multiplet with  $N=8$ supersymmetry has 8 scalars and 8 fermions. 
The scalars are taken to transform in the vector of $SO(8)$. The fermions
transform in the antichiral spinor of $SO(8)$. Supersymmetry generators
transform in the chiral spinor of $SO(8)$. \foot{In previous sections
susy generators transformed in the vector of $SO(8)$. This convention
is related to the convention of this section by a triality transformation.
We must be careful to take this fact into account when applying the results
of the previous sections to this field theory}. The Supersymmetry 
transformation laws are
\eqn\susya{\delta \l^{\at} ={i \over 2} \g^{\mu}\Gamma^a_{\a \at}\epsilon^{\a}
\p_{\mu}\phi^a} 
\eqn\susyb{\delta \phi^{a}={\overline \epsilon}^a\Gamma^a_{\a \at}\l^{\at} }  
Each of the $\phi$ and $\l$ fields is a conformal primary operator 
in the theory on the world volume of an  $M_2$ brane; 
these operators form part of a  
single representation of the $superconformal$ group. The superconformal primary
operators labeling this representations are the $\phi s$; they transform in 
the scalar of $SO(3)$, the vector of $SO(8)$ and have scaling dimension 
$\half$. Making the triality transformation to return to the conventions of 
the previous section, this representation transforms in the scalar of $SO(3)$,
the chiral  spinor of $SO(8)$ and has dimension  $\half$.  This representation
is the `remarkable' or singleton representation of SO(3,2/8), and is 
an example of the isolated $d=3$ representation with $\e=h_1=\half$, whose
existence was predicted in section 4.  

Consider, now, products of $n$ scalar
fields. Such products are also primary operators under the superconformal
algebra, and generate the symmetric tensor product of $n$ singleton 
representations.
These representations are labeled $SO(3)$ scalar;  $SO(8)$ $(n,0,0,..0)$, or 
after triality $SO(8)$ $({n \over 2}, {n \over 2}, ...{n\over 2})$ ; $\e={n \over 2}$. These representations are examples of the isolated $d=3$ 
representations with $\e=h_1={n \over 2}$\foot{These representations have 
been explicitly constructed by the oscillator method in \rgunyadina}. 

	The theory on the world volume of  $m$ coincident $M_2$
branes possesses $m^2$ $N=8$ singleton supermultiplets 
\foot{ a single multiplet in the adjoint of $U(m)$.}. The theory is
 associated with the gauge group 
$U(m)$, and unlike the theory of a single $M_2$ brane is an interacting
superconformal field theory. 
One may form gauge invariant products of $\phi$ fields by taking the 
trace of the products of $\phi$s as $U(m)$ matrices. Such gauge invariant 
products of $\phi$ fields for $n\geq 2$ are primary operators under the 
superconformal group. These operators have the same $SO(3)\times $SO(8)$ \times
D$ labels \rseiberg\ as simple products of $\phi$ fields in the theory of a 
single $M_2$ brane, and therefore generate the same representations of the 
superconformal algebra.

\subsec{{\bf d=4 N=4 SYM- The World Volume theory of the 3 brane.}}

Consider $N=4,d=4$ $U(n)$ Super Yang Mills.  
The microscopic fields of the theory constitute a $U(n)$ adjoint multiplet 
containing
6 real scalars transforming in the vector of $SO(6)$, a vector 
gauge boson which is an $SO(6)$ scalar, 4 complex chiral spinors which are
chiral $SO(6)$ spinors. 

Under supersymmetry these fields transform as \foot{
Spinors are written in 2 component notation. $\l$ denotes a chiral spinor,
and $\lb$ denotes its hermitian conjugate which is antichiral. 
By the product of 2 spinors we mean
$\t\l$ we mean $\t_{\a}\l_{\b}\ep^{\a\b}$ where $\ep^{12}=1$
The R symmetry
has been displayed as $SU(4)$. $ij$ are $SU(4)$ fundamental (vector) indices.}   
\eqn\transa{\d A_{\mu}=i\ep_i\s_{\mu}\lb^i-\l_i\s_{\mu}\eb^i }
\eqn\transb{\d M_{ij}=\ep_i\l_j-\ep_j\l_i+\ep_{ijkl}\eb^k\lb^l }
\eqn\transc{\d\l_i = -\half i\s^{\mu\nu}\ep_i F_{\mu\nu}+
	   2i\s^{\mu}\p_{\mu}M_{ij}\eb +2i[M_{ij},M^{jk}]\ep_k}
  
The microscopic fields of our theory lie in a supermultiplet under the 
superconformal algebra- this is the short multiplet with $j_1=j_2=0$
$\e=1$ and $SO(6)$ rep =(1,0,0)\foot{Note however that the operators in this
representation for $n>1$ are not gauge invariant}.
 The scalar fields $M_{ij}$ are the superconformal primary operators for 
this multiplet - their scaling dimension is the dimension appropriate to 
free scalar fields in  $d=4$.

As in the previous subsection we may identify the trace of 
symmetric products of $p$ $M$ fields as primary superconformal operators that
transform in the representation $j_1=j_2=0$ ; $\e=p$ ; $SO(6)$ rep =(p,0,0), i.e. 
$SU(4)$ Dynkin labels =(0,p,0)\foot{These representations have been
explicitly constructed by the oscillator method in \rgunaydinb }. 

These are
examples of short representations of the $d=4,n=4$ algebra.
To see this note that this $SU(4)$ Dynkin 
labeling corresponds to Young Tableaux with $r_1=r_2=p$ , $r_3=0$, where
$r_i$ is the number of boxes in $i^{th}$ row of the Dynkin diagram. For 
the representation under consideration $d'^1_{44}=d'^3_{41}=p+2$;
$d'^2_{41}=d'^4_{44}=p$. Since $j_1=j_2=0$ and $d^2_{41}=d^4_{44}$ there exist
a special short representation at $\e = d^2_{41}=p$ \foot{See the full listing
of results in the next section}, which is exactly the physical case.

\subsec{{\bf The world volume theory of the $M_5$ brane}}

Consider a $d=6$ $N=2$ superconformal field theory. The only matter 
multiplet has 6 scalars $\phi^a$ transforming as an $SO(6)$ vector, a single self
dual two form field $B_{\mu\nu}$, and 4 chiral spinors $\l$, transforming 
under 
$SO(6)$ as a chiral spinor. The free theory (ie the theory of a single 
$M_5$ brane in the decoupling limit $l_p -->0$) has only one such matter 
multiplet (an explicit action may be found in section 4 of \rkallosh\ for 
instance). The supersymmetry transformation properties of this theory are
given in \rkallosh\ .

Each of the $\phi$ ,$H$ and $\l$ fields in the theory on the world
volume of an  $M_5$ brane is a primary operator of the conformal group; 
these operators form part of a single representation of the superconformal 
algebra. The  primary operators for this representation are the $\phi s$.
The representation labels are $SO(6)$ scalar; $SO(5)$ vector ; $\e=2$.  

	The theory on the world volume of  $m$ coincident $M_5$
branes possesses $m^2$ $N=8$ singleton supermultiplets 
\foot{ a single multiplet in the adjoint of $U(m)$.}. The theory is
 associated with the gauge group 
$U(m)$. Gauge invariant products of $\phi$ fields are
the traces of the products of $\phi$s as $U(m)$ matrices. These 
symmetric products are superconformal primary operators with labels 
$SO(6)$ scalar; $SO(5)$ $(n,0)$ ; $\e=2n$. These are examples of 
scalar representations with $h_1=h_2=h_3=0$, $\e=2 (l_1+l_2)$ , whose existence
we were unsure of in the previous section. Apparently these representations
exist as short representations of the superconformal algebra.

	To establish  the 
existence of the other short representations of the $d=6, n=2$ algebra,
in the next two subsections we review the oscillator construction 
of this algebra given in  \rgunaydinc . 

\subsec{{\bf Oscillator Constructions of Supergroups}}

I now describe the oscillator method for the construction of 
superalgebras and some of
their unitary representations. This method was developed by Gunaydin and 
collaborators (see the references in \rgunyadina\ , 
\rgunaydinb\ ).

Notice that it is extremely easy 
to make an explicit $unitary$ construction of a $U(N)$ Lie algebra using 
free bosonic or fermionic oscillators. The simplest construction of this 
sort takes the form $T^i_j = a^i a_j $. 
\foot{We use the notation of appendix 1. $T^i_j$ is the element
of the complexified lie algebra corresponding to the $N \times N$ matrix with
unity in the $i^th$ row, $j^th$ column, and zeroes everywhere else. Elements
of the real Lie Algebra of $U(N)$ are $T^i_j+T^j_i$ and $i(T^i_j-T^j_i)$.
The upper index = $U(N)$ fundamental; lower index = $U(N)$ antifundamental ; 
$a^i = a_i^{\dagger}$, $a_i$ = bosonic / fermionic oscillator.}

An extension of this construction consists of replacing $a^i a_j$ by
$a^ia_j + b^ib_j +...$ where $b...$ are independent oscillators. 
This extension seems trivial, but it enlarges the space on which we work,
and so permits the construction of more general representations
of the group, as we will see.

To make an oscillator construction of an arbitrary group, one searches for a 
unitary (compact) subgroup; constructs the lie algebra of that unitary
group using oscillators as outlined above, and then attempts to 
construct the lie algebra of 
the rest of the group using the same oscillators. Similarly for Lie supergroups.
We will present an oscillator construction for the (N=2,d=6) superconformal 
algebra in the next subsection. 

\subsec{{\bf The Oscillator Construction  of the d=6, N=2 algebra}}

Much of this subsection is contained in \rgunaydinc , 
and is reworked here for the convenience of the reader.
 
The even part of $SO(6,2 / 2)$ is $SO(6,2) \times SO(5)$. It is 
natural to implement the $SO(3,2)$ with bosonic oscillators and the 
$SO(5)$ with fermionic oscillators. We proceed to do so.

The maximal compact subgroup of $SO(6,2)$ is $SO(6)\times SO(2) =U(4)$. 
Using the philosophy of sec 2.2, this compact part may be
thought of as consisting of Euclidianized Lorentz transformations and 
dilatations : the remainder of the group consists of 
momentum and special conformal transformations. These are vectors under
$SO(6)$ but have weight $\pm 1$ under $U(1)$ dilatations, and hence
are respectively antisymmetric tensors of the fundamental / antifundamental
representations of $U(4)$. The construction of this group with 
$n=2p$ \foot{Construction with an odd number of oscillators turns out not 
to be possible} oscillators is
\eqn\unit{T^i_j= a^i a_j+b_jb^i}
\eqn\mom {S_{ij}= a_i b_j- a_jb_i }
\eqn\speciala {S^{ij}= a^i b^j- a^jb^i } 
Here we have suppressed an internal `which oscillator' index on $a's$ and 
$b's$. There are $p$ varieties of $a$ and $b$ oscillators, a summation over 
the $p$ oscillator flavours is implied in all formulae unless otherwise 
specified.  
We 
identify $D=\half T^i_i$ , Lorentz group = traceless $T^i_j$ , Momentum
$= S^{ij}$ , special conformal $=S_{ij}$.

$SO(5)$ may be constructed in an analogous fashion. Its maximal 
subgroup is $SO(3)\times U(1)=U(2)$. The remaining
generators of $SO(5)$ consist of 2 $SO(3)$ vectors, linear 
combinations of which have  weight $\pm 1$ under the $U(1)$ and so 
are symmetric tensors in the 2 / 2* representation of $U(2)$. 
The construction of the $SO(5)$ algebra is almost identical to the 
construction of $SO(6,2)$, with $i,j$ indices
being replaced by $\mu , \nu$ indices that run from  1-2, bosons
replaced by fermions and antisymmetric tensors replaced by symmetric
tensors
\eqn\unita{M^{\mu}_{\nu}= \a^{\mu} a_{\nu}-b_{\nu}b^{\mu}}
\eqn\moma {A_{\mu \nu}= \a_{\mu} \b_{\nu} + \a_{\nu}\b_{\mu}}
\eqn\momba {A^{\mu \nu}= \a^{\mu} \b^{\nu} + \a^{\nu}\b^{\mu}}

Oscillator representations of $SO(5)$ are easily constructed.
Representations are decomposed into
states of equal weight under $Y=\half M^{\mu}_{\mu}$, which appear in 
multiplets of $SO(3)$. The $Y$ value and the spin $s$ of the $SO(3)$ 
representation 
of the lowest weight state label representations. It is easy to translate
to the more conventional GZ highest weight labeling of $SO(5)$ 
representation : $(l_1,l_2)=(-x , s)$.

The full Lie super algebra SO(6,2/2) may be constructed easily using 
oscillators introduced in the two paragraphs above. Define 2 super tuples of 
oscillators:
$$ A_m =(a_i, \a_{\mu}) ; B_m=(b_i , \b_{\mu} ) $$
The generators of SO(6,2/2) may be constructed with $n=2p$ families of 
oscillators as  
\eqn\unitb{M^m_n= A^m A_n+ (-1)^{deg(m)deg(n)}B_nB^m }
\eqn\momb {S_{mn}= A_m B_n+ A_nB_m }
\eqn\specialb {S^{mn}= A^m B^n+ A^nB^m }
The even part of the superalgebra is $SO(6,2) \times SO(5)$ ; our construction
on the even part reduces to the constructions of these algebras presented 
above. 
Odd elements of the
superalgebra carry a single bosonic ($U(4)$) index, and one fermionic 
($U(2)$) index. Odd elements with raised / lowered bosonic indices are 
identified with $Q's$ /$S's$ , the generators of supersymmetry, and 
super special conformal transformations, in the algebra. 
\foot{Once again this identification is in the sense of sec 2.2 }

Representations of the full supergroup $SO(4,2/2)$ obtained by the oscillator
method are naturally decomposed into states of equal weight under 
$\half M^m_m=D+Y=K$, and appear in multiplets of the supergroup $U(4/2)$. 
Representations are labelled by the $U(2/4)$ multiplet that states with the 
lowest $K$ value 
transform in. Translation into the more usual labeling (of the lowest $D$
value appearing in the representation, and the $SO(6)\times SO(5)$ 
representation that
states with this $D$ value transform in), is easy. 
$U(4/2)$ representations representations consist of a finite number of 
representations of  $U(4) \times U(2)$. The whole representation is 
uniquely specified by the $U(4)\times U(2)$ representation in this group
that has lowest $\e$. With this understanding $U(4/2)$ representations are
labeled by 
$U(4)\times U(2)=U_D(1) \times SO(6) \times U_Y(1)\times SO(3)$ weights.
We want to transform this into a set of $.U_D(1)\times SO(6) \times SO(5)$
labels - that is easily achieved.
$U(1)_D$ and $SO(6)$ labels map into each other:
$U_Y(1) \times SO(3)$ labels map into $SO(5)$ weights according to the 
formulae presented above when analysing representations of $SO(5)$.

Consider some specific representations of this algebra that may be constructed
using this method. 

C1) Representations constructed using $n=2p$ oscillators, and with the
fock space vacuum chosen as lowest weight state, possess a lowest weight
state that transforms as a singlet under $U(4/2)$ , and with weights
$\e=D=2p$; $Y=-p$. The lowest $D$ weight of these states is, therefore, 
$\e=2p$ ; states with this scaling dimension transform as scalars under
$SO(6)$ but in the $(p,0)$ GZ highest weight representation of $SO(5)$.
These are the short representations of this algebra that appear
n the field theory of the $M_5$ brane as explained behind.

C2) Representations constructed using $n=2p$ oscillators, and with 
lowest weight states =$\a_{(1)}^{\mu}|0>$\foot{ (1) is a `which oscillator' index.
Here and in the constructions given  below, one really has in mind the 
following. One constructs a U(4/2) multiplet lowest weight state by acting
on fock space by a set of super oscillator raising operators with 
some symmetry properties between them. One then studies the states produced
and picks out those with the lowest $\e$ value. One could not have,
for instance, considered the state $a^i|0>$ as a lowest weight state 
because that state appears in the
same multiplet as the state $\a^{\mu}|0>$ and the later has lower $\e$ weight.
As a less trivial example of a lowest weight state, consider the state formed
when $n+2$ super oscillators, each of flavour 1, are super symmetrized,  and 
act on the fock space. The lowest weight state has 2 of  the oscillators being
$\a$ mutually antisymmetrized  ( only  2  because you cannot anti symmetrize
more than 2 $\a s$ of a given flavour), and the rest of the oscillators as
$a s$ symmetrized. This is the situation in C5) below. The multiplets to 
which the lowest states in the remaining constructions below belong to 
are : C3) 2 super oscillators of the same flavour, supersymmetrized acting
on fock vacuum. C4)  2 super oscillators of flavour 1 supersymmetrized and 2
super oscillators of flavour 2 supersymmetrized acting on vacuum. 
C6) n+2 oscillators of flavour 1 supersymmetrized and 2 oscillators of 
flavour  2 supersymmetrized acting on fock vacuum }  
 have $\e=2p$ ,are $SO(6)$ scalars
and transform in $SO(5)$ under $(p-\half , \half)$. These are also examples
of short representations with $\e=2(l_1+l_2)$.

	C3) Representations constructed using $n=2p$ oscillators, and with 
lowest weight states =$\a_{(1)}^{\nu}\a_{(1)}^{\mu}|0>$,
have $\e=2p$ ,are $SO(6)$ scalars
and transform in $SO(5)$ under $(p-1, 0)$. These are examples
of short representations with $h_1=h_2=h_3=0$  and $\e=2(l_1+l_2)+2$.

	C4) Representations constructed using $n=2p$ oscillators, and with 
lowest weight states =$\a_{(1)}^{\nu}\a_{(1)}^{\mu}\a_{(2)}^{\phi}
\a_{(2)}^{\theta}|0>$ \foot{ Of course this construction makes sense only 
for $p \geq 2$},
have $\e=2p$ ,are $SO(6)$ scalars
and transform in $SO(5)$ under $(p-2, 0)$. These are examples
of short representations with $h_1=h_2=h_3=0$  and $\e=2(l_1+l_2)+4$.

	C5) Representations constructed using $n=2p$ oscillators, and with 
lowest weight states =$\a_{(1)}^{\nu}\a_{(1)}^{\mu} a_{(1)}^{i_1}
a_{(1)}^{i_2}...a_{(1)}^{i_n}|0>$,
have $\e=2p+{n\over 2}$ ,$SO(6)$ weights $({n\over 2} , {n\over 2},  ... {n\over 2}) $, 
and transform in $SO(5)$ under $(p-1, 0)$. These are examples
of short representations with $h_1=h_2=h_3={n\over 2}$  and 
$\e=h_1+2(l_1+l_2)+2$.

	C6) Representations constructed using $n=2p$ oscillators, and with 
lowest weight states =$\a_{(1)}^{\nu}\a_{(1)}^{\mu}\a_{(2)}^{\phi}
\a_{(2)}^{\theta} a_{(1)}^{i_1}
a_{(1)}^{i_2}...a_{(1)}^{i_n}|0>$ \foot{ Of course this construction makes sense only 
for $p \geq 2$},
have $\e=2p+{n\over 2}$ , have $SO(6)$ weights $({n\over 2} , {n\over 2},  ... {n\over 2}) $ 
and transform in $SO(5)$ under $(p-2, 0)$. These are examples
of short representations with $h_1=h_2=h_3={n\over 2}$  and 
$\e=h_1+2(l_1+l_2)+4$.

In summary, using the oscillator method we have managed to construct all the 
short representations of the $d=6, n=1$ algebra whose existence we were in 
doubt about. Since the $d=6, n=2$ algebra contains a $d=6, n=1$ sub-algebra,
this construction  has established the existence of the corresponding 
representations for $d=6,n=1$ as well. 	
	
\newsec{Results}

\cl{{\bf Representations of the Conformal Algebra }}

Representations of the conformal algebra in spacetime dimension $d$ are 
labeled by
a scaling dimension $\e$ and $SO(d)$ weights $(h_1, h_2 ,..h_{[d/2]})$. 
Necessary conditions for these representations to be unitary are
$$\e \geq {d-2 \over 2}$$ for the scalar representation (all $h$s =0) ; 
$$\e \geq {d-1 \over 2}$$ for the spinor representation.
$$\e \geq  d-1 $$ for the vector representation.
And
$$\e \geq |h_i| +d -i-1$$ for any other representation R, where $i$ is the 
smallest value s.t. $h_i \geq |h_{i+1}|+1$, if there exists such an $i$, or is
equal to $[d/2]$ ([ ] takes integer part), if there is no such $i$.

These conditions are also sufficient to guarantee unitarity in $d=3,4,$ according
to \revans, \rmack . I have not attempted to find sufficient conditions in 
arbitrary dimension.

Free representations of the conformal algebra, in the form of a local operator 
obeying a free wave equation, exist only when\foot{this result was obtained 
largely in \rsiegel .}

a) d=even. $h_1=h_2=...=|h_{[d/2]}|=h$. In that case $\e = {d-2\over 2}+h$.

b) d=odd. $h_1=h_2=...=h_{[d/2]}=h$ and $h=0$ or $h=\half$. In that case 
$\e= {d-2 \over 2} +h$

\cl{{\bf Unitary Representations of Superconformal Algebras }}

\cl{{\bf d=3}}

Representations are labeled by and $SO(3)$ weight j, and by $h_1 ... h_{n/2}$, 
$SO(n)$ highest weights and a lowest dimension $\epsilon_0$. We report results
for even n.

If j $> 0$ then the condition 
$\epsilon_0 \geq j+1+h_1$ is necessary and sufficient to
guarantee unitarity. A representation that saturates this equality in  $\epsilon_0$ is in a short representation of the Superconformal Algebra, and so the 
scaling dimension of the operator is likely to remain constant as one turns
interactions on or off in the theory.

If j = 0 then unitary representations exist for $\e=h_1$ and for 
$\e \geq h_1 +1$. The representation at the isolated value of $\e$,
 and the one at the value of $\e$ that saturates the inequality above, are
short.

If n=1 then $\e \geq \half$  if $j=0$ and $\e \geq j+1$ otherwise.

\cl{{\bf d=4}}

These results were first obtained by Dobrev and Petkova in \rdpresult\ .
Representations are labeled by 2 $SU(2)$ - Lorentz - $j$s, 
by a $U(1)$ R charge\foot{For reasons mentioned in Section 4 the case n=4 
is special. In this case the simple LSA has no R charge in it - the R symmetry
group is $SU(n)$. The results stated here all hold on setting R=0 in all 
formulae} and by an $SU(n)$ representation - which we label by 
positive integers $r_k$, the number of boxes in the $k^{th}$ row of the 
Young Tableaux. 

If $j_1 \neq 0 \neq j_2$ then $\epsilon_0 \geq max(d'^3_{nn} , d'^1_{n1})$
where $d'$s are defined in eqs \done\ - \dfour . These conditions are necessary
and sufficient to guarantee unitarity. A representation with dimension 
saturating the inequality is a short representation.

If $j_1$ = 0, $j_2 \neq 0$ and $d'^3_{nn}> d'^2_{n1}$ then the condition 
above continues to be necessary and sufficient to guarantee unitarity.  

If $j_2$ = 0, $j_1 \neq 0$ and $d'^4_{nn}<  d^1_{n1}$ then the condition 
above continues to be necessary and sufficient to guarantee unitarity.  

If $j_1$ = 0, $j_2 \geq 0$ and $d'^3_{nn}\le d'^2_{n1}$ then 
$\e \geq d'^1_{n1}$ is sufficient to guarantee unitarity. It is not however n
ecessary -
there exists a (short representation) at $\e = d'^2_{n1}$ that is 
unitary. There are no other exceptions.  

If $j_2$ = 0, $j_1 \geq 0$ and $d'^4_{nn}\ge d'^1_{n1}$ then $\e \geq d'^3_{nn}$
is  sufficient to guarantee unitarity. It is not however necessary -
there exists a (short representation) at $\e = d'^4_{nn}$ that is 
unitary.

If $j_1=j_2=0$, then unitary representations exist for $\e \geq 
max(d'^3_{nn}, d'^1_{n1})$ and for $\e=d'^2_{n1}$ if $d'^2_{n1} \geq d'^3_{nn}$
and for  $\e=d'^4_{nn}$ if $d'^4_{nn} \geq d'^1_{n1}$
and for  $\e=d'^2_{n1}$ if $d'^2_{n1} = d'^4_{nn}$.
Unitary representations exist for no other values of $\e$, except for the 
vacuum representation dealt with below.  

Finally if all labels $j_1, j_2, r_i, R $ are zero then there exists in 
addition to the representations above, a trivial one dimensional representation.

\cl{{\bf d=5}}

Representations are labeled by scaling dimension $\epsilon_0$, by an R symmetry
($SU(2)$) half integer $k$, and by $S0(5)$ (Lorentz) weights [$h_1, h_2$].

If $h_1 \neq 0 \neq h_2 $ then $\epsilon_0 \geq h_1 +h_2 +4 +3k$ forms the 
necessary and sufficient conditions for the occurrance of unitary 
representations. The representation at the saturating dimension is short.

If $h_1 \neq 0$, $h_2 = 0$ then there do
exist unitary representations for $\e \geq h_1 +3k+4$. The only other 
dimension at which a unitary representation may exist is $\e = h_1+3k +3$.
My calculations suggest, but do not prove, that a unitary representation
exists at this isolated value of $\e$. The representation at the dimension 
that saturates the inequality, and at the isolated dimension (if it exist)
are short representations.

If $h_1  =h_2=0$ then unitary representations exist for $\e \geq 3k+4$. The 
only other values of $\e$ at which unitary representations could possibly exist
are $\e = 3k$ and $\e = 3k+3$. My calculations suggest, but do not prove,
that unitary representations do indeed exist at these isolated values of $\e$ .
The representation at the value of $\e$ that saturates the inequality, and at
those at the isolated values of $\e$, (if they exist), are short 
representations. 

\cl{{\bf d=6}}

Supersymmetry generators are taken to be chiral spinors, i.e., ($h_1, h_2, h_3$)=
$(\half , \half , \half)$.

\cl{{\bf n=1}}
 
Representations are labeled by scaling dimension $\epsilon_0$, by an $SU(2)$
R symmetry
half integer $k$, and by $S0(6)$ (Lorentz) weights [$h_1, h_2, h_3$].

a) $h_2 -{1\over 2} \geq |h_3+{1\over 2}|$
Unitary representations exist if and only if $\e \geq h_1 + h_2 -h_3 +4k+ 6$
The representations at the value of $\e$ that saturates this inequality is 
short.

b) $h_1-h_2 \geq 1$ and the condition of part a) is not met.
Unitary representations do exist when $\e \geq h_1 + h_2 -h_3 +4k+ 6$ . The
representation at the dimension that saturates this bound is short. The only
other value of $\e$ at which unitary representations may exist are
$\e = h_1 - h_2 +h_3 + 4k +4$. My calculations suggest, but do not prove the 
existence of a unitary representation at this dimension. If this representation
exists, it is short.

c) $h_1=h_2=h_3=h \neq 0$
Unitary representations do exist for $\e \geq 6+h+4k$. The representation
at $\e=6+h+4k$ is short. Short unitary
representations also exist for $\e=2+h+4k,\  \  4+h+4k$. Unitary 
representations may or may not exist for $\e$ in the interval 
$(2+h+4k,4+h+4k)$\foot{ I am tempted to guess that representations exist 
through the interval}.

d) $h_1= h_2=h_3=0$
Unitary representations do exist for $\e \geq 4k+6$. The representation 
at $\e=4k+6$ is short. Short unitary
representations also exist for $\e=4k,\  \ 4k+2, \  \ 4k+4$. Unitary 
representations may or may not exist for $\e$ in the interval $(4k+2, 4k+4)$
\foot{I am tempted to guess that representations exist throughout 
the interval}.

\cl{{\bf n=2}}

Representations are labeled by scaling dimension $\epsilon_0$, by an R symmetry
$(SO(5)$) half weights $l_1, l_2$, and by $S0(6)$ (Lorentz) weights 
[$h_1, h_2, h_3$].

Our results are exactly those for n=1, with k being replaced by 
${(l_1 + l_2)\over 2}$.

\cl{{\bf Acknowledgements}}
I would like to acknowledge useful discussions with almost every high energy
theory student at the physics department at Princeton, especially R.Gopakumar,
M.Krogh, S.Lee and A.Mikhailov (who suggested the use of characters in 
appendix 2). I would like to thank R.Gopakumar and M.Rangamani for reading 
a preliminary draft of this paper and suggesting improvements.
I would like to thank V.Dobrev for responding promptly and 
lucidly with clarifications on his work, and for drawing my attention to 
\rdpreview . I am grateful to M. Gunaydin for a communication that drew 
my attention to his work on the oscillator representations of superalgebras. 
A conversation with K. Intrilligator  and another with C.Callan
helped clarify some aspects
related to section 2. Finally, I would like especially to acknowledge the large
contribution of N.Seiberg to this paper, in suggesting the problem, and in 
providing guidance and advice during every stage of its completion.

\newsec{Appendix 1: $U(n)$ and $SU(n)$}

We specify here the conventions we use when dealing with the R symmetry
$U(n)$ in the case $d=4$.

Consider the matrices $(T_{ij})_{pq} = \delta_{ip}\delta_{jq}$ . 
Complex linear combinations of these matrices $T_{ij}+T_{ji}$  and $(-i)(
T_{ij}-T_{ji})$ are hermitian, and so form a basis for the defining representation of the Lie Algebra
of $U(n)$. We will find it convenient to work with the T matrices themselves
and their images in every $U(n)$ representation, rather then their Hermitian 
linear combinations $T_{ij}$. 
They obey commutation relations.
\eqn\tcom{[T_{ij}, T_{mn}] = T_{in}\delta_{jm} - T_{jm}\delta_{in} }
Note $ T_{ij}^{\dagger} = T_{ji} $
If $Q_{i}$ is an operator that transforms in the defining representation
 of $U(n)$, 
then it obeys
\eqn\tqcom{[T_{ij}, Q_{m}] = \delta_{jm}Q_i}
If $S_{i}$ transforms in the antifundamental representation of $U(n)$, then
\eqn\tscom{[T_{ij}, S_{m}] = -\delta_{im}S_{j} }
Therefore the $i$ in $T_{ij}$ is a fundamental index, the $j$ an 
antifundamental index.

The various weights of the fundamental representation are defined as $ \nu '_i 
(i=1,...,n)$. If a state $\psi$ has weight $\Sigma a_i\nu'_i$ then 
$T_{mm} \psi =a_m \psi$ (no sum over m). $T_{mm}$, thus, are natural 
Cartan Generators for $U(n)$. 

Define $R = \Sigma T_{mm} $. R is a generator for the $U(1)$ sub-algebra of 
$U(n)$. In the defining representation it is the identity matrix. All 
vectors in the defining representation have R charge 1. All vectors in 
the antifundamental representation have R charge -1. All Ts commute with
R, i.e., have zero R charge.

Consider the decomposition of $U(n)$ into $U(1) \times SU(n)$. The $U(1)$
generator is R above. $SU(n)$ generators may be chosen to be all off diagonal
$Ts$, and $T'_{m}= T_{mm} - R/n$ (no sum over m). The n $T's$ clearly obey
the constraint $\Sigma T'_{i} = 0$. Therefore we regard $T'_{i}$ for $i = 1..n-1$
as the Cartan generators of our $SU(n)$ algebra. Let $\nu_{i} = \nu_{i}'$
restricted in their action to the Cartan generators of $SU(n)$ rather than 
$U(n)$. 
Then the arbitrary $SU(n)$ weight vector may be written as 
$\sum_{i=1}^{n-1} r_i \nu_i=\sum_{i=1}^{n-1} R_i \mu_i$ 
where we have defined $ \mu_i = \sum_{j=1}^i \nu_j $. 
Here $r_i = R_i - R_{i-1} $. $\mu_j$s are fundamental weights for $SU(n)$. 
Sums are taken only to $n-1$ because of the identity $\sum_{i=1}^n \nu_i = 0.$
 When the highest
weight of a representation is written in this form, $R_i$ refers to the
number of columns of length $i$ in the Young Tableaux for that representation, $r_i$ refers to the number of boxes in the ith row of the Tableaux.
 
Acting on a state with weights specified by {$r_i$}, our $T'$ operators
yield $T'_m \psi = (r_m - {\Sigma r_i \over n} ) \psi $ We define the ordering
convention $\nu_1 \ge \nu_2 \ge \nu_3 ...\geq  \nu_{n-1}$ $T_{ij}$ are raising
and lowering operators, with roots $\nu_i - \nu_j$. Therefore $T_{ij}$
is a raising operator for $i \le j$ and a lowering operator otherwise.

I define a $U(n)$ Casimir of a representation by 
$C(R)=\sum_{i,j=1}^n (T_{ij}T_{ji}).$ This may now be computed as a 
function of the highest wts of the $SU(n)$ part of the representation, and 
of the R charge of the representation (by acting the casimir on the
highest weight state, and noting that all nonzero elements in the sum
may be replaced by commutators). The answer is
$C(R)= \sum_{i=1}^n (r_i - \Sigma(r_j)/n)(r_i -\Sigma(r_j)/n - 2i) + 
{R^2 \over n}$
For the purposes of the formula above, we define $r_n$ = 0.
As a check on the formula, we note that in the case of $SU(2)$ $T_{ij}T_{ji}$ =
2( $J^2$) + $R^2$/2, and so independently we expect the casimir to be given
by $2(j)(j+1) + R^2/2$ . Setting $r_1=2j$, $r_2=0$ in the formula above, that
is exactly what we get. Again, the Casimir for the fundamental representation 
should be n (from definition), again this is reproduced by our formula.

The similarly defined Casimir for $SU(n)$ is obtained by setting R=0 in the 
formula above.

\newsec{ Appendix 2: Decomposition of $Vector \times R$ and $Spinor \times R$ for $SO(N)$.}

In this appendix we will derive the irreducible representation content of 
Vector $\times$ (Rep) and Spinor $\times$ rep in $SO(N)$.

Define the character of a representation to be
\eqn\character{\chi (\{h_j\}, t_j) = Tr(\exp(t_iH_i)) } in that representation
where $\{h_i\}$ are the highest weights labeling the representation, $t_i$ are
real variables, and $H_i$ are the Cartan Generators of our group in the standard
basis. The trace above is taken in the representation labeled by $\{h_i\}$. We
sum over $i$ in the RHS of the formula above. Since the character of a product representation
must be equal to the product of its factor characters, and must 
also be equal to 
the sum of the characters of the its irreducible representation content, the
character is a useful aid to performing a Clebsch Gordan 
decomposition.

\cl{{\bf SO(2n+1)}}

The character of the vector representation is clearly given by $1 +\sum_{i=1}^n 2Cosh(t_i)$. The character of the spinor is given by 1+$\Pi_{i}2Cosh(t_i)$.
The product of the general representation is given by the famous Weyl formula
\eqn\chara{ \chi (h_i, t_i)= {det( \sinh[t_i(h_j+(n-j)+\half)]) \over
det( \sinh[t_i( (n-j) +\half)])}}
The product of the character of an arbitrary 
representation with that of the vector representation, leaves the denominator
in the \chara\ unchanged, but turns the numerator into the sum of 
2n+1 determinants, one being the original determinant and the others being 
 obtained from the original determinant by taking $h_i$
to $h_i + \llap{-} 1$ for any given i=j, leaving $h_i$ unchanged for i $\neq $ j.
Note that for any rep $h_1 \geq h_2 \geq h_3 ...\geq h_n\geq 0$. Of the 2n determinants
we obtain by the process described above not all will be nonzero - only those
whose new $h_i$ values are ordered as above will yield a nonzero determinant.
For instance if originally $h_1 = h_2$, and you form a new determinant with 
$h_2$ augmented by unity, then two rows of the matrix whose determinant is
being considered become identical, and the determinant is zero. Similarly, if
you form a new determinant with $h_1$ decreased by unity you get zero.

Thus we obtain a simple rule : $\{h_i\} \times (vector)$ = $\{h_i\}$ + All reps obtained 
from $\{h_i\}$ by adding or subtracting unity from a single h, provided that
the set $\{h_i\}$ thus obtained continues to obey 
$h_1 \geq h_2 \geq h_3\geq ...\geq h_n\geq 0$.

The rule for the spinor is derived as easily, and I state it below.
 : $\{h_i\} \times (spinor)$ = All reps obtained 
from $\{h_i\}$ by either adding or subtracting 1/2 from every $h_i$, provided that
the set $\{h_i\}$ thus obtained continues to obey 
$h_1 \geq h_2 \geq h_3 ...\geq h_n\geq 0$.  

This rule could lead to a maximum of $2^n$ representations.

\cl{{\bf SO(2n)}}

A condition $SO(2n)$ highest weights must obey is  
$h_1 \geq h_2 \geq h_3 ...|h_n|\geq 0$. (recall
$h_n$ may be negative).
The Weyl formula for $SO(2n)$ is 
\eqn\charc{ \chi (h_i, t_i)= {det( \sinh[t_i(h_j+n-j)]) + det( \cosh[t_i(h_j+n-j)])\over det( \sinh[t_i( n-j)])}}
Where the character is defined as for $SO(2n+1)$.
The vector and 2 spinor characters may be worked out from the formula above. 
Redoing our $SO(2n+1)$ computation leads to the following rules.

: $\{h_i\} \times (vector)$ =  All reps obtained 
from $\{h_i\}$ by adding or subtracting unity from a single h, provided that
the set $\{h_i\}$ thus obtained continues to obey 
$h_1 \geq h_2 \geq h_3\geq ...\geq |h_n|\geq 0$.
 
 : $\{h_i\} \times \pm chirality\  \ spinor$ = All reps obtained 
from $\{h_i\}$ by either adding or subtracting $\half$ from every $h_i$, the number of subtractions being (even/odd), provided that
the set $\{h_i\}$ thus obtained continues to obey 
$h_1 \geq h_2 \geq h_3 \geq ...\geq |h_n|\geq 0$. 
 
\listrefs

\end

\end